\documentclass[12pt]{article}
\usepackage{epsf}
\usepackage{cite}
\usepackage{amssymb}
\def\g{\gamma}
\def\s{\sigma}
\def\th{\theta}
\def\pp{\pi^{+}}
\def\pn{\pi^{0}}
\def\i{\prime}
\def\thq{\th_{\g\g^{\i}}^{cm}}
\def\thg{\th_{\g\g^{\i}}}

\def\thP{\th_{\g^{\i}\pp}}
\def\a{\alpha}
\def\b{\beta}
\def\O{\Omega}
\def\m{\mu}
\def\mm{m_{\pi}}
\def\n{\nu}
\def\f{\varphi}
\def\d{\Delta}
\def\e{\varepsilon}

\def\ab{(\a-\b)_{\pi^{\pm}}}

\def\gp{\g p \to \g \pp n}
\def\gn{\g p \to \pn \pp n}
\def\EQ{\begin{equation}}
\def\EN{\end{equation}}
\def\be{\begin{eqnarray}}
\def\en{\end{eqnarray}}
\def\unit{\times 10^{-4} {\rm fm}^3}
\def\Unit{/10^{-4} {\rm fm}^3}

\begin{document}

\begin{center}
{\Large\bf Measurement of the $\pp$-meson polarizabilities via
the $\gp$ reaction}
\end{center}
\vskip 0.5cm

\begin{center}
{\large
J.~Ahrens$^{1}$,
V.M.~Alexeev$^{2}$,
J.R.M.~Annand$^{3}$,
H.J.~Arends$^{1}$,
R.~Beck$^{1}$,
G.~Caselotti$^{1}$,
S.N.~Cherepnya$^{2}$,
D.~Drechsel$^{1}$,
L.V.~Fil'kov$^{2}$\footnote{E-mail: filkov@sci.lebedev.ru},
K.~F\"ohl$^{4}$,
I.~Giller$^{5}$,
P.~Grabmayr$^{6}$,
T.~Hehl$^{6}$,
D.~Hornidge$^{7}$,
V.L.~Kashevarov$^{2}$,
M.~Kotulla$^{8}$,
D.~Krambrich$^{1}$,
B.~Krusche$^{8}$,
M.~Lang$^{1}$,
J.C.~McGeorge$^{3}$,
I.J.D.~MacGregor$^{3}$,
V.~Metag$^{9}$,
M.~Moinester$^{5}$,
R.~Novotny$^{9}$,
M.~Pfeiffer$^{9}$,
M.~Rost$^{1}$,
S.~Schadmand$^{9}$,
S.~Scherer$^{1}$,
A.~Thomas$^{1}$,
C.~Unkmeir$^{1}$,
and Th.~Walcher$^{1}$}
\end{center}
\vskip 0.5cm

\noindent
1 Institut f\"ur Kernphysik der Johannes-Gutenberg-Universit\"at, Mainz,
Germany \\
2 P.N.Lebedev Physical Institute, Moscow, Russia \\
3 Department of Physics and Astronomy, Glasgow University, Glasgow, UK \\
4 School of Physics, University of Edinburg, Edinburg, UK \\
5 School of Physics and Astronomy, Tel Aviv University, Tel Aviv, Israel \\
6 Physikalisches Institut, Universit\"at T\"ubingen, Germany \\
7 Mount Allison University, Sackville, NB, Canada \\
8 Institut f\"ur Physik, Basel, Switzerland \\
9 II. Physikalisches Institut, Universit\"at Giessen, Giessen, Germany \\
\vskip 1cm

\begin{center}
{\bf Abstract}
\end{center}
\vskip 0.2cm

An experiment on the radiative $\pp$-meson photoproduction from the proton
($\gp$) was carried out at the Mainz Microtron MAMI in the
kinematic region
537~MeV~$<E_{\g}<817$~MeV, $140^{\circ}\le\thq\le 180^{\circ}$.
The $\pp$-meson polarizabilities have been determined
from a comparison of the data with the predictions of two different
theoretical models, the first one being based on an effective pole model
with pseudoscalar coupling while the second one is based on diagrams
describing both resonant and nonresonant contributions. The validity of
the models has been verified by comparing the predictions with the
present experimental data in the kinematic region where the pion
polarizability contribution is negligible ($s_1<5\mm^2$) and where
the difference between the predictions of the two models
does not exceed 3\%. In the region, where the pion
polarizability contribution is substantial ($5<s_1/\mm^2<15$,
$-12<t/\mm^2<-2$), the difference $(\a-\b)_{\pp}$
of the electric ($\alpha$) and the magnetic ($\beta$) polarizabilities
has been determined. As a result we find:
$(\a-\b)_{\pp}=(11.6\pm 1.5_{stat}\pm 3.0_{syst}\pm 0.5_{mod}) \unit$.
This result is at variance with recent calculations in the framework
of chiral perturbation theory.


\vskip 1cm

\section{Introduction}

The pion polarizabilities characterize the deformation of the pion
in an external electromagnetic field. The values of
the electric ($\a$) and magnetic ($\b$) polarizabilities
depend on the rigidity of the  composite particle and provide important
information of the internal structure.
Very different values for the pion polarizabilities have been predicted
in the past. All predictions agree, however, that the sum of the two
polarizabilities of the $\pi^{\pm}$-meson is very small. On the other hand,
the value of the difference of the
polarizabilities is very sensitive to the theoretical models.
The investigations within the framework of
chiral perturbation theory (ChPT) predict $(\a-\b)_{\pi^{\pm}}\approx
5.4$ \cite{don,bellucci} at one-loop  and $4.4\pm 1.0$ at
two-loop order
\cite{burg}. Note that here and in the following the polarizabilities are
given in units of
$10^{-4}$fm$^3$. The calculation in the extended Nambu-Jona-Lasino model
with a linear realization of chiral $U(3)\times U(3)$ symmetry
\cite{ivan1} results in $\a_{\pi^{\pm}}=-\b_{\pi^{\pm}}=3.0\pm 0.6$.
The application of dispersion sum rules (DSR) at a fixed value of the
Mandelstam variable $u=\mm^2$ \cite{rad,kash} leads to
$(\a-\b)_{\pi^{\pm}}=10.3\pm 1.9$ and $(\a-\b)_{\pn}=-3.01\pm 2.06$.
DSR at finite energy \cite{petr} gave a similar
result for the charged pion ($(\a-\b)_{\pi^{\pm}}=10.6$) and a smaller
value with large uncertainties for the neutral pion,
$(\a-\b)_{\pn}=0.3\pm 5$. A calculation using
the linear $\s$ model with quarks and vector mesons to one loop
order predicts $(\a-\b)_{\pi^{\pm}}=20$ \cite{bern},
and the Dubna quark confinement model \cite{ivan} results in
$(\a-\b)_{\pi^{\pm}}=7.05$ and $(\a-\b)_{\pn}=1.05$.

The experimental information available so far for the polarizability
of the pion is summarized in table 1.
The scattering of high energy pions off the Coulomb
field of heavy nuclei \cite{antip} resulted in
$\a_{\pi^{-}}=-\b_{\pi^{-}}=6.8\pm 1.4\pm 1.2$ assuming
$(\a+\b)_{\pi^-}=0$.
This value agrees with the prediction of DSR but is
about 2.5 times larger than the ChPT result. The experiment of the
Lebedev Institute on radiative pion photoproduction from the proton
\cite{lebed} has given $\a_{\pp}=20\pm 12$. This value has large error
bars and shows the largest discrepancy with regard to the
ChPT predictions. The attempts to determine the polarizability from the
reaction $\g\g\to\pi\pi$ suffer greatly from theoretical \cite{pen} and
experimental \cite{boy} uncertainties.
The analysis of MARK II and Crystal Ball data in ref. \cite{holst} finds
no evidence for a violation of the ChPT predictions. However, even
changes of polarizabilities by $100\%$ and more are still compatible
with the present error bars. As seen from table 1,
our present experimental knowledge about the pion polarizability
is still quite unsatisfactory.

The present work is devoted
to the investigation of the radiative $\pp$-meson photoproduction from the
proton with the aim to determine the $\pp$-meson polarizability. The
experiment on this process has been carried out at the Mainz Microtron MAMI.

The content of this paper is as follows.
The connection of pion polarizability with Compton scattering
on the pion and radiative photoproduction of the $\pp$-meson and
calculations of the cross section for this process are given in sect.~2.
The experimental setup is described in sect.~3. The analysis of the
experimental data and the determination of the
$\pp$-meson polarizability are given in sect.~4.
The discussion of the results obtained is in sect.~5. Conclusions are
presented in sect.~6. The details of the calculations of baryon resonances
and the description of the orthogonal amplitudes method are given in
Appendices A, B, and C.

\section{Radiative pion photoproduction from the proton and
the pion polarizability (theory)}
\subsection{ Compton Scattering on the pion and pion
polarizabilities}

Expanding the Compton scattering amplitude on the pion with respect to the
photon energy and taking  account of terms up to second order, we have
\cite{petr2,klein,jon}
\EQ
T_{\g\pi \to\g\pi} = T_{B} +
8\pi\mm\omega_{1}\omega_{2}
[\vec{\e}_{2}\cdot\vec{\e}_{1}\a_{\pi} +
(\vec{\kappa}_{2}\times\vec{\e}_{2})\cdot(\vec{\kappa}_{1}\times\vec{\e}_
{1})\b_{\pi}] + \cdots ,
\EN
where $\vec{\e}_{1}(\vec{\e}_{2})$, $\omega_{1}(\omega_{2})$
and $\vec{\kappa}_{1}(\vec{\kappa}_{2})$ are the polarization vector,
the energy
and the direction of the initial (final) photon, respectively,
$\mm$ is the $\pi^{\pm}$ meson mass, and $T_{B}$ is the Born amplitude.
The low energy expansion for the helicity amplitudes $M_{++}$ and $M_{+-}$
can be written as \cite{fil1}:
\be
M_{++}& = &M_{++}^{({\rm B})} +
2\pi\mm (\a - \b)_{\pi^{\pm}}, \nonumber \\
M_{+-}& = &M_{+-}^{({\rm B})} +
\frac{2\pi}{\mm}(\a + \b)_{\pi^{\pm}}.
\en

The low energy expressions for the
differential and total cross sections for Compton scattering on
charged pions take the form \cite{fil1,drec}
\be\label{com}
\frac{d\s_{\g\pi}}{d\O} & = &
\left( \frac{d\s_{\g\pi}}{d\O} \right)_{\rm B} -
\frac{e^{2}}{4\pi} \frac{\mm^{3}(s_{1}-\mm^{2})^{2}}
{4s_{1}^{2}[(s_{1}+\mm^{2})+(s_{1}-\mm^{2})z]}\nonumber\\
  & &\times\left\{ (1-z)^{2}(\a_{\pi^{\pm}}-\b_{\pi^{\pm}}) +
\frac{s_{1}^{2}}{\mm^{4}}(1+z)^{2}(\a_{\pi^{\pm}}+
\b_{\pi^{\pm}}) \right\},
\en
\be
\s_{\g\pi} & = & \s_{\rm B} -
\frac{e^{2}}{4\pi}\frac{2\pi\mm}{s_{1}}
\left\{\left[\frac{(s_{1}-\mm^{2})^{3}-
6s_{1}\mm^{4}+2\mm^{6}}{2\mm^{2}s_{1}} \right.\right.\nonumber\\
&&+\left.\left. \frac{2\mm^{2}s_{1}}{s_{1}-
\mm^{2}}\ln\left(\frac{s_{1}}{\mm^{2}}\right) \right]\a_{\pi^{\pm}} +
\frac{(s_{1}-\mm^{2})^{3}}{2\mm^{2}s_{1}}\b_{\pi^{\pm}} \right\}
\en
where $z=\cos \thq$, the index $B$ indicates the Born
cross sections and $s_1$ is the square of the total energy
in the center-of-mass system (c.m.s.) for the $\g\pi\to\g\pi$
reaction (see eq. (\ref{var1})).

The differential Born cross sections is given by
\EQ
\left(\frac{d\s_{\g\pi}}{d\O}\right)_{\rm B} = \frac12\left(\frac{e^2}{4\pi}
\right)^2
\frac{1}{s_1}\left[1+\left(\frac{(s_1-\mm^2)+(s_1+\mm^2)z}{(s_1+\mm^2)+
(s_1-\mm^2)z}\right)^2\right].
\EN

We work at values of $s_1$ up to $15\mm^2$. It has been shown in ref.
\cite{kash}
that the contributions of the scalar-isoscalar two-pion correlations
($\s$ meson) are noticeable at such high values of
$s_1$.
On the other hand, the contribution of other mesonic
resonances $(\rho,\; a_1,\; b_1,\; a_2)$
is negligible in this region.
The $\s$ meson was considered in ref. \cite{kash} as an effective description
of the strong $S$-wave $\pi\pi$ interaction using the broad Breit-Wigner
resonance expression.

Therefore, we will take account of the $\s$
meson by using the dispersion relation from \cite{kash}
\EQ\label{compt}
M_{++}=M_{++}^{({\rm B})} +2\pi\mm (\a-\b)_{\pi^{\pm}} +\frac{t_1}{\pi}
P\!\!\int\limits_{4\mm^2}^{\infty}\frac{Im M_{++}^{\s}(t_1^{\i})dt_1^{\i}}
{t_1^{\i}(t_1^{\i}-t_1)},
\EN
where
$$
Im M_{++}^{\s}(t_1)=\frac{g_{\s}\Gamma}{(M^2_{\s}-t_1)^2+\Gamma^2},
$$
$$
g_{\s}=8\pi\frac{M_{\s}+\sqrt{t_1}}{\sqrt{t_1}}\left(\frac{\frac23
\Gamma_{\s}\Gamma_{\s\to\g\g}}{M_{\s}\sqrt{M_{\s}^2-4\mm^2}}
\right)^{\frac12},
$$
$$
\Gamma=\frac{\Gamma_{\s}}{2}(\sqrt{t_1}+M_{\s})\left(\frac{t_1-4\mm^2}
{M^2_{\s}-4\mm^2}\right)^{\frac12},
$$
and $t_1$ is the momentum transfer of the process $\g\pi\to\g\pi$.

The parameters of the $\s$ meson have been determined in
\cite{kash} from a fit of the experimental data for the process
$\g\g\to\pn\pn$:
\EQ\label{sigma}
M_{\s}=547\pm 45\; {\rm MeV}, \quad \Gamma_{\s}=1204\pm 362\; {\rm MeV}, \quad
\Gamma_{\s\to\g\g}=0.62\pm 0.19\; {\rm keV}.
\EN

\subsection{Kinematics of radiative $\pp$-meson
photoproduction from the proton.}

Radiative pion photoproduction from the proton is described by
five independent kinematical invariants, which are expressed in the
laboratory system:

\be\label{var1}
s & = &(p_{1}+k_{1})^{2}=m_p^{2}+2m_pE_{\g}, \nonumber \\
t & = &(p_{2}-p_{1})^{2}=(m_n-m_p)^2-2m_p(E_n-m_n), \nonumber  \\
t_{1} & = &(k_{2}-k_{1})^{2}=-2E_{\g}E_{\g^{\i}}(1-\cos \thg), \\
s_{1} & = &(k_{2}+q_{2})^{2}=\mm^{2}+2E_{\g^{\i}}(q_{20}-
\mid \vec{q}_{2}\mid\cos\thP),  \nonumber \\
s_{2} & = &(p_{2}+q_{2})^{2}=s+t_{1}-2m_pE_{\g^{\i}}, \nonumber
\en
where $E_{\g}$ and $E_{\g^{\i}}$
are the energies of the initial and final photons, respectively, $m_n$
is the mass and $E_n$ is the energy of the neutron,
$q_{20}$ and $\vec{q}_{2}$ are the energy and the momentum of
the pion, and $m_p$ is the proton mass (see fig.~1).

The variable $s_2$ can be expressed in terms of the so-called Treiman-Yang
angle $\f_b$ as \cite{kaj}
\be\label{s2}
&&s_2=s-\frac{1}{(s_1-t)^2}\left\{2\sqrt{F_1F_2}\cos\f_b+(s_1-t)\times\right.\\
\nonumber
&&\left.\left[(s_1-\mm^2)(s-m_p^2)-t_1(s-m_n^2+s_1)\right]+
2t_1s_1(s-m_p^2)\right\},
\en
where
\EQ\label{f1}
F_1=t_1\left[s_1t_1+(s_1-\mm^2)(s_1-t)\right],
\EN
\EQ\label{f2}
F_2=t\left[s(s-m_p^2-m^2_n-s_1+t)+m_p^2(m_n^2-s_1)\right]+s_1\left[s_1 m_p^2+
(s-m_p^2)(m_n^2-m_p^2)\right]
\EN
and
\EQ\label{fib}
\cos\f_b=\frac{(\vec k_1\times\vec p_1)\cdot(\vec k_1\times \vec q_2)}
{|\vec k_1\times\vec p_1||\vec k_1\times\vec q_2|}.
\EN
The Treiman-Yang angle $\f_b$ is defined as the angle
 between the planes formed by the momenta
$\vec k_1$, $\vec p_1$ and $\vec k_1$, $\vec q_2$
in the c.m.s. of the $\g\pi$ scattering. The conditions $F_1\le 0$ and
$F_2\le 0$
determine the physical region of the process under investigation.

The pion polarizability can be extracted from experimental data on
radiative pion photoproduction, either by  extrapolating these data to the
pion pole \cite{drec,fil2,fil3,walch1},
or by comparing the experimental cross section with the predictions of
different theoretical models.
The extrapolation method was first suggested in \cite{chew} and has been widely
used for the determination of cross sections and phase shifts of
elastic
$\pi\pi$-scattering from the reaction \mbox{$\pi N\rightarrow \pi \pi N$}.
For investigations of $\g\pp$-scattering this method was first used in
\cite{ksf,lebed}.

However, in order to obtain a reliable value of the pion
polarizability, it is necessary to obtain the experimental data on pion
radiative photoproduction with small errors over a sufficiently
wide region of $t$, in particular, very close to $t=0$ \cite{ahr,ahr1,unk}.

It should be noted that there is an essential difference in
extrapolating the data of the processes $\pi N\to \pi\pi N$ and
$\g p\to \g\pi N$. In the former case, the pion pole
amplitude gives the main contribution in a certain energy region. This
permits to constrain the extrapolation function to be zero
at $t=0$ providing a precise determination of the amplitude. In the
case of radiative pion photoproduction, the pion pole amplitude
alone is not gauge invariant and we must take into account all pion and
nucleon pole amplitudes. However, the sum of these amplitudes does not
vanish at $t=0$. This complicates
the extrapolation procedure by increasing the requirements on the
accuracy of the experimental data.

As the accuracy of the present data is not sufficient for a reliable
extrapolation, the values of the pion polarizabilities have been obtained
from a fit of the cross section calculated by different theoretical
models to the data.

\subsection{ Calculations of the cross section for the \\
reaction $\gp$}

The theoretical calculations of the cross section for the reaction
$\gp$ show that the contribution of
nucleon resonances is suppressed for photons
scattered backward in the c.m.s. of the reaction $\g\pi\to\g\pi$.
Moreover, integration over $\f$ and $\thq$ essentially decreases the
contribution of resonances from the crossed channels.
On the other hand, the difference \mbox{$(\a-\b)_{\pp}$} gives the biggest
contribution to the cross section
for $\thq$ in the region of $140^{\circ}-180^{\circ}$.
Therefore, we will consider the cross section of radiative pion
photoproduction integrated over $\f$ from $0^{\circ}$ to $360^{\circ}$
and over $\thq$ from $140^{\circ}$ to $180^{\circ}$,
\EQ
\int_0^{360^{\circ}}d\varphi\int_{-1}^{-0.766}d\cos\thq\;
\frac{d\s_{\gp}}{dtds_1d\O_{\g\g}},
\EN
where the angle $\f$ is equal to the angle $\f_b$ in (\ref{fib}).

The cross section of the process $\gp$ has been
calculated in the framework of two different models.
In the first model (model-1) the contribution of all
the pion and nucleon pole diagrams is taken into account
using pseudoscalar pion-nucleon coupling (fig.~2) \cite{unk}.

In the second model (model-2), we include the nucleon and the
pion pole diagrams without the anomalous magnetic moments of the nucleons
(fig.~3a-e), and in addition
the contributions of the resonances $\d (1232)$, $P_{11}(1440)$,
$D_{13}(1520)$, and $S_{11}(1535)$ according to fig.~3f.
The amplitude of the $\g\pp$ elastic scattering in this model is
described by eq.~(\ref{compt}).
We will determine the pion polarizability by comparing the experimental data
with predictions of these theoretical models in different regions of $s_1$.
Therefore, we limited ourselves to describing the baryon resonances
by the diagram fig.~3f only, because this diagram has
a pole at $s_1=\mm^2$. The details of the calculations of
these resonance contributions are given in the Appendices.

It should be noted that the contribution of the sum of the pion
polarizabilities is very small in the considered region of
$140^{\circ} \lesssim \theta^{cm}_{\g\g`} \lesssim 180^{\circ}$ .
The estimate shows that the contribution of $(\a+\b)_{\pi^{\pm}}=0.4$ to
the value of $(\a-\b)_{\pi^{\pm}}$ is less than 1\%.

\section{Experimental setup}

The experiment has been performed at the continuous-wave electron accelerator
MAMI B \cite{mami1,mami2} using Glasgow-Mainz photon tagging facility
\cite{ant, hall}. The quasi-monochromatic photon beam covered an energy
range from 537 to 819~MeV with an intensity $\sim 0.6\times 10^6\gamma/$s in
the tagger channel with a 2.3 MeV wide bite for the lowest photon energy.
The average energy resolution was 2~MeV.
The tagged photons entered  a scattering chamber, containing a 3~cm
diameter and 11.4~cm long liquid hydrogen target with Kapton windows.
The target was aligned along the photon beam direction.
The emitted photon $\g^{\i}$, the $\pp$-meson, and the neutron were
detected in coincidence. The experimental setup is shown in fig.~4.

The photons were detected by the spectrometer TAPS \cite{nov,gab},
assembled in a special configuration (fig.~5).
The TAPS spectrometer consists of 528 BaF$_2$ crystals, each
250~mm long (corresponding to 12 radiation lengths) and hexagonally
shaped with an inner
diameter of 59~mm. All crystals were arranged into three blocks. Two
blocks (A,B) consisted of 192 crystals arranged in 11 columns and the third
block (C) had 144 crystals arranged in 11 columns. These three blocks
were located in the horizontal plane around the target at central angles
$68^\circ$, $124^\circ$, $180^\circ$ with respect to the beam axis.
Their distances to the target center were 55~cm, 50~cm and 55~cm,
respectively. All BaF$_2$ modules were equipped with 5~mm thick plastic
veto detectors for the identification of charged particles.

The neutrons were detected by a wide aperture time-of-flight spectrometer (TOF)
\cite{tof1}. It consisted of 111 scintillation-detector bars of dimensions
$50\times200\times3000$~mm$^3$ and 16 counters
($10\times230\times3000$~mm$^3$) which were used as charged-particles
veto detectors. The bars are made from NE110 plastic scintillator and each
bar is read out at each end by a $3''$ phototube type XP2312B. All bars
were assembled in planes 8 deep, each plane having 16 detectors (fig.~4).
This block of
plastic scintillators and  detected neutrons in the energy
region 10-100 MeV with efficiencies varying between $30$ to $50\%$
depending on energy.
The neutron energy determination was by time of flight and for the present
energy range and path the FWHM resolution was $\sim 10\%$.
Accuracy of the horizontal position of TOF was $\pm 10$ cm and the FWHM
resolution of its vertical position was 29 cm \cite{dirk}.


In order to detect the $\pp$-meson two two-coordinate multi-wire proportional
chambers (MWPC) and a forward scintillator detector (FSD) have been developed
and constructed (fig.~6), which provided a fast trigger signal.
The MWPCs were located at $0^{\circ}$ with respect to the beam direction.
Their sensitive areas covered angles in the laboratory system
$\theta\cong2^\circ - 20^\circ$,
$\varphi\cong 0^\circ - 360^\circ$.
The MWPCs have the following characteristics:
\begin{itemize}    
\item sensitive region:  $292\times292$~mm$^2$,
\item photon-beam aperture:  $40\times40$~mm$^2$,
\item anode wires:  gold plated tungsten $20~\m$m,
\item distance between wires:  2~mm,
\item cathode planes:  $25~\m$m aluminum foil,
\item gas windows:  $50~\m$m mylar foil,
\item anode-cathode gap: 5~mm,
\item maximum current for all wires: $100~\m$A.
\end{itemize}
Each MWPC has two perpendicular planes each with 128 wires.

The MWPCs operated with a gas mixture of Argon ($60\%$) and Isobutane ($30\%$).
The gas mixture was blown through the chambers at a flow rate of
180~ml/min. MWPCs were read by LeCroy 2735~DC cards.
The FSD had 16, $1\times2\times30$~cm$^3$ plastic scintillator strips
with a $4\times4$~cm$^2$ hole in the middle for the photon beam. Each
strip was read out by a single photomultiplier tube.

The MWPCs were optimized for high count rates and good efficiency,
described as follows. The measurement of the efficiency was carried
out with a $^{90}$Sr beta source. The MWPC was placed between two
$190\times20\times5$~mm$^3$ plastic scintillators. This detector
arrangement was irradiated by a radioactive source, viewed through a
two millimeter collimator. Thresholds of the discriminators for the
plastic detectors were fixed at 30~mV.
Double coincidence rates between the plastic detectors and triple
coincidences which included the MWPC were measured simultaneously at
different thresholds of the LeCroy 2735~DC cards, for different values
of the MWPC high voltage. A high voltage for MWPC of 3.25~kV with 3~V
threshold for the LeCroy~2735~DC cards was chosen where the maximum
efficiency was about 98~\% and the noise was suppressed.

In order to optimize the MWPC and FSD for high counting rates, a test
measurement
was carried out with untagged bremsstrahlung produced at the tagger
radiator ($E_e=855$~MeV).
The first MWPC was positioned at a distance of 131~cm from the end of the
vacuum beam pipe. A mylar target of 3~mm thickness was placed at a distance
of 45~cm from the foil covering the first chamber. Thresholds of the FSD
discriminators were set to $500\pm 50$~keV on the basis of the Compton
electron response to $^{60}$Co $\g$ rays. The test run results are
presented in table~2. We found that at the maximum intensity in
the tagging system (beam current $\sim150$~nA) the current of the MWPC was
$76~\m$A, well within operational limits.

In the experiment on the radiative pion photoproduction
the first MWPC was centered at a polar angle of $0^{\circ}$ with respect to
the beam axis, and at a distance of 46~cm from the center of the target
to the first wire plane. The second MWPC was placed $\sim 10$ cm
behind the first one and was rotated $45^{\circ}$ in azimuth with the respect
to the front chamber. The FSD was positioned between the first and second
MWPCs. After the commissioning of this experiment in the A2 Tagger Hall at
Mainz, triple coincidence of $\g$, $\pp$, and $n$
were taken over a period of 1150 hours.

\section{Determination of the $\pp$-meson polarizability}
\subsection{Analysis of the experimental data}

In order to determine an efficiency of the $\g$ and $\pp$ detectors
and to check the normalization of the experimental data,
$\pn$-meson photoproduction from the proton has been measured
and the obtained cross section was compared with the well-known values
\cite{mami}.

Neutral pions produced in the liquid hydrogen target were detected via
their two-photon decay with the TAPS spectrometer. The MWPC and FSD were
used to detect the protons, triggered by a coincidence between TAPS
and
FSD. About $2.5\times10^6$ raw events were collected and after kinematic
selection $\sim5\times10^5$ events remained. The background of multiple
pion production was removed by reconstructing the missing mass spectrum
for the reaction $\g p\to \pn X$. Random coincidences were subtracted by
samples selected  outside the prompt coincidence time window with
the photon tagger.

The cross section was obtained using the detector yields, the detection
efficiency of each detector arm, the thickness
of the liquid hydrogen target, and the intensity of the photon beam.
The photon intensity was determined by counting the electrons detected in
the tagging spectrometer focal plane detector.
The tagging efficiency, i.e. the probability of a bremsstrahlung photon passing
through the collimator giving an $e^-$ hit in the focal plane, was
determined by comparing the number of electrons detected in the tagger to
the number of photons detected in a 100\%
efficient lead glass detector, which was moved into the photon beam during
special runs at very low beam intensity.
The angle and energy dependent detection
efficiency of TAPS and FSD+MWPC was calculated by Monte Carlo using GEANT3
\cite{geant}, in which all relevant properties of the setup are taken into
account. As result we obtained experimental data for the
process $\g p\to\pn p$ for the angles $140^{\circ}$, $150^{\circ}$,
$160^{\circ}$, and  $170^{\circ}$ in the energy region $290-810$~MeV.

The angular distribution data in the energy region
$480-530$~MeV are shown in fig.~7. The filled circles
are the data of the present work. The open circles are the data
obtained in \cite{mami}. The results of the theoretical models MAID and
DMT are depicted by the solid and dashed curves, respectively.
The dotted lines are the results of the partial wave analysis SAID of
the world data.

The present angular distributions are in a good agreement
with those of ref. \cite{mami} and with the predictions of
MAID, DMT and SAID for incoming photon energy up to
$\sim 650$~MeV. This result shows that the efficiencies of the $\g$ and
$\pp$-meson detectors correspond well to the simulation result.

To determine the efficiency of the TOF detector, GEANT was improved
by the STANTON program package \cite{stant}, which allowed one to
consider more correctly an interaction of the low energy neutrons
with the plastic scintillator and to convert the ionization losses
of different charged particles into the electron equivalent.
The accuracy of this simulation is limited by the uncertainty of the
cross sections of the elastic scattering of the neutrons with the
protons in the scintillators of 2\%. This has to be added to the
inaccuracy of GEANT for the efficiency calculation of the whole setup.
In ref.~\cite{mami} a similar setup has been investigated with GEANT
and the results have been compared to the accurately measured
$p(\gamma,\pi^0)p$ reaction giving an uncertainty of 3\%. Adding these
two contributions quadratically results in 4\% for the systematic
error of the overall efficiency of the experimental setup.

In order to check the functioning of the neutron detector, data
on double pion photoproduction $\gn$ were selected
and analyzed. To this aim, the $\pn$-meson was reconstructed via the
invariant mass of the two decay photons and then, using the reconstructed
momentum of the neutron, the missing mass spectrum for the $\pp$-meson was
constructed. Alternatively, the invariant mass of the $\pn$ and neutron
gave a prominent peak, the width and the position of which corresponded to
the $\d^0(1232)$ resonance. The analysis of the data obtained for the
process $\gn$ indicated that it proceeds mainly through excitation of the
$\d^0(1232)$ resonance in our kinematical region in agreement with
the finding of ref.~\cite{lang}. The investigation of
this process showed that the selection of the neutrons and determination
of their parameters was correct.

The process $\gp$ was detected by counting the triple coincidences of
the emitted photon, the $\pp$-meson, and the neutron.
Coincident pulses from TAPS and FSD were used as
a pre-trigger. This signal was a "stop" for the tagger focal plane
and a "start" for TAPS and TOF Time Digital Convertors (TDC). Slightly
later coincidence information from the tagger and TOF then determined
if the event was stored or cleared.

The main source of spurious correlated contributions to $\gp$
comes from the $\gn$ reaction.
This background was simulated using the total cross section
measured at MAMI \cite{mainz} and was suppressed by using conservation
of energy and momentum. To this end we compared in the analysis the
invariant variables $t$, $s_1$ and $t_1$ with the variables
$t^{\i}$, $s_1^{\i}$, and $t_1^{\i}$ determined from
the measured data by two different methods \cite{cas}. The values of
$t$ were determined according to eq.~(\ref{var1}) when the neutron
energy was measured by the TOF detector. In the case of $t^{\i}$ this
energy was calculated from the equations of the energy and momentum
conservation and $t^{\i} = f_1(E_{\g},E_{\g^{\i}},\theta_{\g\g^{\i}},
\theta_{\g\pp},\theta_{\g^{\i}\pp})$. The variable $t_1$ was evaluated as
function of $E_{\g}$, $E_{\g^{\i}}$, and $\theta_{\g\g^{\i}}$.
For $t_1^{\i}$ the energy and the angle of the final photon were
expressed through other measured variables and
$t_1^{\i}=f_2(E_{\g},E_n,\theta_{\g n},\theta_{\g\pp},
\theta_{\pp n})$. In the case of $s_1$ and $s_1^{\i}$ we had
$s_1=f_3(E_{\g},E_n,\theta_{\g n})$ and
$s_1^{\i}=f_4(E_{\g},E_{\g^{\i}},\theta_{\g\g^{\i}},
\theta_{\g\pp},\theta_{\g^{\i}\pp})$.

The constraint that
$(t-t^{\i})<0.5\mm^2$, $(s_1-s_1^{\i})<0.5\mm^2$,
$(t_1-t_1^{\i})<0.7\mm^2$ was applied.
Such a kinematic cut, which was chosen on the basis of the
simulation, suppressed
the background to the $3\%$ level relative to the $\gp$ process.

Prompt and random coincidence regions were determined from
a two dimensional triple-coincidence spectrum (fig. 8)
where the time difference between the initial $\g$ and the final $\g^{\i}$
photons $(t_{\g}-t_{\g^{\i}})$, detected by the tagger and TAPS,
and the difference between the final $\g^{\i}$ photon and the
$\pp$-meson $(t_{\g^{\i}}-t_{\pp})$, detected by TAPS and FSD,
are displayed at the horizontal axes.
The sharp peak in this figure corresponds to the triple coincidence of
the reaction under study. The ridge along the
$(t_{\g}-t_{\g^{\i}})$ axis and the plateau
are the contributions of random double coincidences for
the initial $\g$ and final $\g^{\i}$ photons and for TAPS--FSD,
respectively. The yield was calculated by subtracting the numbers of the
random coincidences for two photons and TAPS--FSD
with appropriate weights given by the relative width of the cuts, from
the number of the events under the sharp peak.

A missing-mass spectrum of the reaction $\g p\to\g n X$, constructed
using the photon and the neutron momenta measured in this experiment,
is shown in fig.~9. This spectrum shows a peak at $M_X=139.7$~MeV
with a resolution $\s=20$~MeV corresponding to the $\pp$-meson. It agrees
well with the simulation results. As seen from this figure, a clean
separation of radiative \mbox{$\pp$-meson} photoproduction on
the proton from the competing background reaction $\gn$ is obtained.

As result we have identified about $4\times10^5$ radiative \mbox{$\pp$-meson} 
photoproduction events in the kinematic region:
$E_{\g}=537 - 817 MeV$, $1.5\mm^2\le s_1\le 15\mm^2$,
$-12\mm^2<t<t_{max}$, $140^{\circ}\le\thq<180^{\circ}$.

\subsection{Derivation of the polarizability}

In order to reduce the influence of
the nucleon resonances, the differential cross section was
integrated over the angles $\thq$ from $140^{\circ}$ to $180^{\circ}$ and
$\f_b$ from $0^{\circ}$ to $360^{\circ}$.

To increase our confidence that
the model dependence of the result is under control we limited ourselves
to kinematic regions where the difference between
model-1 and model-2 did not exceed $3\%$ when $(\a-\b)_{\pp}$ is
constrained to zero.
First, we consider the kinematic region where the contribution of the
pion polarizability is negligible, i.e. the region
$1.5\mm^2\le s_1<5\mm^2$.

In fig.~10, the experimental data for the differential
cross section, averaged over the full photon beam energy interval from
537 MeV up to 817 MeV and over $s_1$ in the indicated interval,
are compared to predictions of model-1 (dashed
curve) and model-2 (solid curve). The dotted
curve is the fit of the experimental data in the region of
$-10\mm^2<t<-2\mm^2$.
As seen from this figure, the theoretical curves are very close to the
experimental data.
This means that the dependence of the differential cross section on the
square of the four momentum transfer $t$ which is basically the kinetic
energy of the neutron (see eq.~(\ref{var1})) is well reproduced by using
the mentioned GEANT simulations for the efficiency. On the other
side, there is a small difference of less than 10\% in the absolute
overall efficiency which could be due to a theoretical, an
experimental, or both deviation from the true cross section.
However, since we are only interested in the change of the
curves with $\ab$ we slightly adjusted the experimental cross
section to the theoretical.

Then we investigated the kinematic region where the polarizability
contribution is biggest. This is the region $5\mm^2\le s_1<15\mm^2$ and
$-12\mm^2<t<-2\mm^2$.
In the range $t>-2\mm^2$ the polarizability contribution is small
and also the efficiency of the TOF is not well known here. Therefore,
we have excluded this region.

In the considered region of the phase space, with maximum
sensitivity of the cross section to the polarizability but small
differences between the considered theoretical models, we obtained
the cross section of the process $\gp$ integrated over
$s_1$ and $t$.  All events are divided into 12 bins of the initial
photon energy. For each bin $i$, the cross section
$\s^i$ is calculated from
\EQ
\s^{i} = Y^{i}/\e^{i} N^{i}_{\g} N_{t},
\EN
where
 $Y^i$ is the number of the selected events after background
subtractions, $\e^i$ the detection efficiency for the $\gp$ channel,
$N_t$ the number of protons per area in the 11.4 cm of the $LH_2$ target,
and $N^{i}_{\g}$ the number of photons passing through the target
in the same time interval as for the integration of $Y^i$.

The cross sections are calculated
according to model-1 and model-2 for two different values of
$(\a-\b)$ within the phase space covered by the experiment.
The obtained experimental cross sections and their theoretical
predictions for $(\a-\b)_{\pp}=0$ and $14 \unit$  are presented in fig.~11.
The error bars are the quadratic sum of statistical and systematic errors.

The systematic error is due to the uncertainties of the time
and kinematic cuts ($\pm 1\%$ for each cut), the number of target
protons ($\pm 1.5\%$), the photon flux ($\pm 2\%$),
and the detection efficiency calculations ($\pm 4\%$). As a result,
we get the limiting systematic deviation for the cross section
of $\pm 5\%$. This is equivalent to a rectangular error distribution
with a $\pm 5\%$ limit. The root-mean square error of this distribution
is then $\sigma_{syst} = 5/\sqrt{3}\% \simeq 3\%$.

Comparing these experimental data
with predictions of the models we find values of $(\a-\b)^i_{\pp}$
and corresponding errors $\d_{stat}(\a-\b)^i_{\pp}$ and
$\d_{syst}(\a-\b)^i_{\pp}$ for each experimental point $i$.

As an example of such a procedure, fig. 12 shows the experimental
value of the cross section at $E_{\g}=653$ MeV with
statistical and systematic errors
and the dependence of the cross section on
$(\a-\b)_{\pp}$ calculated in the framework of model-1. Comparing
these experimental data with the model predictions we obtain for this
energy bin
\EQ
(\a-\b)_{\pp}=9.03^{+3.58(stat) +3.12(syst)}_{-3.28(stat) -2.89(syst)}.
\EN
This result has slightly
asymmetric errors, but making the approximation that they are
symmetrical and taking their average values,
the value of $(\a-\b)_{\pp}$ is obtained by averaging over all 12 bins
of $E_{\g}$ as
\EQ
(\a-\b)_{\pp}=\frac{\sum_i (\a-\b)^i_{\pp}w_i}{\sum_i w_i},
\EN
where $w_i=1/(\d_{stat}(\a-\b)^i_{\pp})^2$.

The statistical error for the averaged value $(\a-\b)_{\pp}$ is calculated
as
\EQ
\d_{stat}(\a-\b)_{\pp}=(\sum_i w_i)^{-1/2}.
\EN
The systematic errors are correlated for the cross sections at the
different values of the energy $E_{\gamma}$.
They give
contributions to the systematic errors for $(\a-\b)^i_{\pp}$ at
different $i$ having different statistical weights.
Therefore, the systematic error $\d_{syst}(\a-\b)_{\pp}$
is determined by calculating a weighted average as follows:
\EQ
\d_{syst}(\a-\b)_{\pp}=\frac{\sum_i \d_{syst}(\a-\b)^i_{\pp}w_i}
{\sum_i w_i}.
\EN

Using this procedure separately for each model, we obtain:
\EQ\label{kr}
(\a-\b)_{\pp}=(12.2\pm 1.6_{stat}\pm 3.3_{syst})\unit \qquad {\rm(model-1)},
\EN
\EQ\label{f}
(\a-\b)_{\pp}=(11.1\pm 1.4_{stat}\pm 2.8_{syst})\unit \qquad {\rm(model-2)}.
\EN

As indicated above, the dominant systematic error is caused by the
uncertainties in the neutron detector efficiency. However, this is
a difficult problem to overcome because better neutron detectors are
hardly possible.

An additional independent analysis \cite{igor} of the experimental data
was carried out by a constrained $\chi^2$ fit \cite{silin}.
The reaction $\gp$ is identified by 15 quantities for the momenta of
all five participating particles. In this experiment 14 quantities were
measured for each event, the complete 4-momenta of all particle except
the pion for which only the direction angles were measured.
Considering the over-determined situation, a constrained $\chi^2$ fit
was employed to determine how well the measured event detection angles
$(\theta,\,\varphi)$ agreed with the measured momenta magnitudes $P$,
and as well to give the optimum $(\theta,\,\varphi,\,P)$ values for
each event for all five participating particles. To assure convergence,
the variations were done only for the magnitudes of the final state
momenta. All combinations of measured particles were tested to see which best
satisfy the $\gp$ reaction.
This method allows us for each event to find the combination
of measured particles which best fit the $\gp$ reaction kinematics.
The $P$ values best fitting $(\theta,\,\varphi)$ correspond
to the minimum of the constructed constrained $\chi^2$ functional.

Simulation studies of the constrained fit method showed that the
described reconstruction algorithm based on the kinematic fit
$\chi^2$ selection criteria can be successfully used for the
suppression of the double pion photoproduction background and
reconstruction of the $\gp$ reaction.

A series of seven analyses with $\chi^2<4,\,5,\,6,\,7,\,8,\,9,\,10$
has been performed. The value for $(\a-\b)_{\pp}$ stabilizes for
$\chi^2<5$ and the values for $\chi^2<4$:
\EQ\label{ig1}
(\a-\b)_{\pp}=(10.1\pm 2.6_{stat}\pm 3.0_{syst})\unit \qquad
{\rm(model-1)},
\EN
\EQ\label{ig2}
(\a-\b)_{\pp}=(10.3\pm 2.3_{stat}\pm 2.7_{syst})\unit \qquad
{\rm(model-2)}
\EN
agree very well with the first analysis giving it additional support.
However the statistical
accuracy of this method is less significant and, therefore, we give
the result of the optimized cut analysis (\ref{kr}), (\ref{f}),
averaged for model-1 and model-2:
\EQ\label{ab}
(\a-\b)_{\pp}=(11.6\pm 1.5_{stat}\pm 3.0_{syst}\pm 0.5_{model})\unit.
\EN

\section{Discussion}

   The experimental result (\ref{ab}) for the difference $(\a-\b)_{\pp}$
of the electric and magnetic polarizabilities provides an important piece of
information about the hadronic structure of the pion as tested with soft
external electromagnetic fields.
   Moreover, from a theoretical point of view, there is another reason
for the extraordinary interest in and importance of a precise experimental
determination of the charged-pion polarizabilities.
   The approximate $\mbox{SU}(2)_L\times\mbox{SU(2)}_R\times\mbox{U}(1)_V$
chiral symmetry in the two-flavor sector of QCD results in a Ward identity
which relates Compton scattering on a charged pion,
$\gamma\pi^+\to\gamma\pi^+$,
to radiative charged-pion beta decay, $\pi^+\to e^+\nu_e\gamma$.
   The corresponding low-energy theorem was originally derived by
Terentev \cite{terent} in the framework of the partially
conserved axial-vector current (PCAC) hypothesis in combination with
current algebra.
   This PCAC prediction is equivalent to the result obtained using
chiral perturbation theory at leading non-trivial order
(${\cal O}(p^4)$) and can be written in the form
\begin{equation}
\label{alphachptop4}
\alpha_{\pi^+}=-\beta_{\pi^+}=2 \frac{e^2}{4\pi} \frac{1}{(4\pi F_\pi)^2
\mu}\frac{\bar l_6-\bar l_5}{6},
\end{equation}
where $F_\pi=92.4$ MeV is the pion decay constant
and $(\bar l_6-\bar l_5)$ is a linear combination of scale-independent
parameters of the Gasser and Leutwyler Lagrangian \cite{gasser}.
   At lowest non-trivial order $({\cal O}(p^4))$ this difference is related to
the ratio $\gamma=F_A/F_V$ of the pion axial-vector form factor $F_A$
and the vector form factor $F_V$ of radiative pion beta decay
\cite{gasser}:
\begin{displaymath}
\gamma=\frac{1}{6}(\bar l_6-\bar l_5).
\end{displaymath}
   Once this ratio is known, chiral symmetry makes an {\em absolute}
prediction for the  polarizabilities.
   This situation is similar to the case of $\pi\pi$ scattering
\cite{weinberg}, where the $s$-wave $\pi\pi$-scattering lengths are
predicted once $F_\pi$ has been determined from pion decay.
   Using the most recent determination $\gamma=0.443\pm 0.015$ by the PIBETA
Collaboration  \cite{frlez} (assuming $F_V=0.0259$ obtained
from the conserved vector current hypothesis) results in the
${\cal O}(p^4)$ prediction
\begin{displaymath}
\alpha_{\pi^+}=(2.64\pm 0.09) \times 10^{-4}\, \mbox{fm}^3,
\end{displaymath}
where the estimate of the error is only the one due to the error of $\gamma$
and does not include effects from higher orders in the quark mass
expansion.
   Clearly, there will be corrections to the prediction of
eq. (\ref{alphachptop4}).
   The results of a two-loop analysis (${\cal O}(p^6)$) of the charged-pion
polarizabilities have been worked out
in ref. \cite{burg}\footnote[2]{Ref. \cite{burg} uses
$(\bar l_6-\bar l_5)=2.7\pm
0.4$ instead of $2.64\pm 0.72$ which was obtained in
ref. \cite{gasser} from $\gamma=0.44\pm 0.12$.
Correspondingly, this also generates a smaller error in the
${\cal O}(p^4)$ prediction $\alpha_{\pi^+}=(2.7\pm 0.4) \times 10^{-4}\,
\mbox{fm}^3$ instead of $(2.62\pm 0.71)\times 10^{-4}\,
\mbox{fm}^3$.}:
\begin{eqnarray}
\label{alphaplusbetachptop6}
(\alpha+\beta)_{\pi^+}&=&(0.3 \pm 0.1) \times 10^{-4}\, \mbox{fm}^3,\\
\label{alphaminusbetachptop6}
(\alpha-\beta)_{\pi^+}&=&(4.4\pm 1.0) \times 10^{-4}\, \mbox{fm}^3.
\end{eqnarray}
   First of all, we note that the degeneracy $\alpha=-\beta$ has been
removed at ${\cal O}(p^6)$.
   The corresponding corrections amount to an 11\% (22\%) change of
the ${\cal O}(p^4)$ result for $\alpha_{\pi^+}$
($\beta_{\pi^+}$), indicating a similar rate of convergence as for
the $\pi\pi$-scattering lengths \cite{gasser,bijnens}.
   The effect of the new low-energy constants appearing at ${\cal O}(p^6)$
on the pion polarizability was estimated via resonance saturation by
including vector- and axial-vector mesons.
   The contribution was found to be about 50\% of the two-loop result.
   However, one has to keep in mind that ref. \cite{burg} could
not yet make use of the improved analysis of radiative pion decay which,
in the meantime, has also been evaluated at two-loop accuracy
\cite{Bijnens:1996wm,Geng:2003mt}.
   Taking higher orders in the quark mass expansion into account,
Bijnens and Talavera obtain $(\bar l_6-\bar l_5)=2.98\pm
0.33$ \cite{Bijnens:1996wm}, which would slightly modify the leading-order
prediction to $\alpha_{\pi^+}=(2.96\pm 0.33)\times 10^{-4}\,
\mbox{fm}^3$ instead of $\alpha_{\pi^+}=(2.7\pm 0.4) \times 10^{-4}\,
\mbox{fm}^3$ used in ref. \cite{burg}.
   Accordingly, the difference $(\alpha-\beta)_{\pi^+}$ of eq.
(\ref{alphaminusbetachptop6}) would increase to
$4.9\times 10^{-4}\,\mbox{fm}^3$ instead of $4.4\times 10^{-4}\,
\mbox{fm}^3$, whereas the sum would remain the same as in
eq.\ (\ref{alphaplusbetachptop6}).
   A value of $4.9\times 10^{-4}\,\mbox{fm}^3$ deviates by 2
standard deviations from the experimental result of eq. (\ref{ab}).
   Nevertheless, both the precision measurement of radiative pion
beta decay \cite{frlez} and of radiative pion photoproduction
indicate that further theoretical and experimental work is needed.
In particular, the analysis of ref. \cite{frlez} suggests
an inadequacy of the present $V-A$ description of
the radiative beta decay, which would also reflect itself
in an inadequacy of the ChPT description in its present form.

   A different approach for obtaining a theoretical prediction for
the difference $(\a-\b)$ is the application of dispersion sum rules
(DSR).
   In ref. \cite{rad} dispersion relations at a fixed value of the
Mandelstam variable $u=\m^2$ without subtraction were applied
to the helicity amplitude $M_{++}$ of elastic $\g\pi$ scattering:
\EQ\label{dsr}
(\a-\b)=\frac{1}{2\pi^2\m}\left\{\int\limits_{4\m^2}^{\infty}
\frac{Im M_{++}(t^{\i},u=\m^2)\,dt^{\i}}{t^{\i}}+
\int\limits_{4\m^2}^{\infty}\frac{Im M_{++}(s^{\prime},u=\m^2)\,
ds^{\i}}{s^{\i}-\m^2}\right\}.
\EN
   The biggest contribution to this DSR is given by the strong
$s$-wave $\pi\pi$ interaction in the $t$ channel.
   This interaction can be effectively described by the $\s$ meson using
a broad Breit-Wigner resonance expression.
   The parameters of such a $\s$ meson have been determined in
ref. \cite{kash} from a fit to the experimental data for the
$\g\g\to\pn\pn$ process (see eq. (\ref{sigma})).
   A saturation of the DSR (\ref{dsr}) by the $\rho(770)$, $b_1(1235)$,
$a_1(1260)$, and $a_2(1320)$ mesons in the $s$ channel and the $\s$,
and $f_0(980)$ mesons in the $t$ channel leads to
\cite{kash}
\be\label{piplus}
 &&\quad \rho \qquad a_1 \qquad b_1 \quad \; a_2 \quad\;\; f_0 \qquad \s
\nonumber \\
(\a-\b)^{DSR}_{\pi^{+}}&=&-1.2+2.1+0.9-1.4+0.4+9.5=(10.3\pm 1.9)
\unit,
\en
   where the error indicated for this value is caused by the error for the
$\s$ meson parameters.
   This value of $(\a-\b)_{\pi^{+}}$ is in
agreement with the experimental result of eq. (\ref{ab}) but differs
significantly
from the ChPT result of eq. (\ref{alphaminusbetachptop6}).
   On the other hand, the DSR for $(\a-\b)_{\pn}$ yields \cite{kash}
\be\label{pimin}
 &&\quad\; \rho \qquad\;\; \omega \qquad\;\; \f \qquad\; f_0
\qquad\;\;\; \s
\nonumber \\
(\a-\b)^{DSR}_{\pn}&=&-1.79-11.69-0.04+0.44+10.07=(-3.01\pm 2.06)
\unit
\en
which, within the errors, is not in conflict with the
two-loop ChPT predictions
$\alpha_{\pi^0}=(-0.35\pm 0.10)\times 10^{-4}\,\mbox{fm}^3$ and
$\beta_{\pi^0}=(1.50\pm 0.20)\times 10^{-4}\,\mbox{fm}^3$
\cite{bellucci}
and with the experimental values $(\a-\b)_{\pn}=(-1.6\pm 2.2)\unit$ and
$(-0.6\pm1.8)\unit$ obtained in refs. \cite{kash} and
\cite{ser}, respectively.

One might ask for explanations of the difference between ChPT and our
experiment without questioning the validity of ChPT. The most
obvious idea is to consider the "off-shellness" of the initial pion.
Clearly, this issue cannot be addressed consistently as an independent
effect \cite{scher,fear} and would require a more sophisticated
analysis such as, e.g., a full ChPT calculation at the one-loop
level which is beyond the scope of the present work. However,
in order to obtain an estimate of this effect,
we have calculated $(\a-\b)_{\pi^+}$ with
the help of dispersion relations for a mass of the initial pion equal
to $t$ and found that such a correction is less than 5\%.

\section{Conclusion}

A measurement of radiative $\pp$ photoproduction from the proton
($\gp$) was carried out at the Mainz tagged photon facility
in the kinematic region
537~MeV~$<E_{\g}<~817$~MeV, $140^{\circ}~\le\thq\le~180^{\circ}$.
The difference of the electric and magnetic $\pp$-meson polarizabilities
has been determined from a comparison of the
experimental data with predictions of two different theoretical models
which included or neglected baryon resonances. In order to
reduce the contribution of the baryon resonances, the differential cross
section was integrated over $\thq$ from $140^{\circ}$ to $180^{\circ}$
and over $\f$ from $0^{\circ}$ to $360^{\circ}$.
To further reduce the model dependence, the kinematic region was
chosen such that the difference between the predictions of the two
considered models did not exceed $3\%$ for $(\a-\b)_{\pp}=0$.

The experimental detection efficiency was normalized by comparing
the theoretical predictions with the experimental differential cross
sections in the kinematic region where the
pion polarizability contribution is negligible $(1.5<s_1/\mm^2<5)$.

In the region, where the pion polarizability contribution is substantial
($5<s_1/\mm^2<15$, $-12<t/\mm^2<-2$), the difference of the electric and
magnetic $\pp$-meson polarizabilities was determined by the comparison
of the experimental data for the cross section of radiative pion
photoproduction with the predictions of two theoretical models under
consideration. As a result we have obtained:
$(\a-\b)_{\pp}=(11.6\pm 1.5_{stat}\pm 3.0_{syst}\pm 0.5_{model}) \unit$.
This result is consisted with the result of ref.~\cite{antip}
and at variance with recent calculations in the framework
of chiral perturbation theory.

\section*{Acknowledgments}
The authors thank the accelerator group of MAMI as well as many other
scientists and technicians of the Institut f\"ur Kernphysik at the
University of Mainz for their excellent help.
This work was supported by Deutsche Forschungsgemeinschaft (SFB 201 and
443), Russian Foundation for Basic Research (grant No 030204018),
the UK Engineering and Physical Sciences Council,
the Swiss National Science Foundation,
and in part by the Israel Science
Foundation founded by the Israel Academy of Sciences and Humanities.

\renewcommand{\theequation}{A\arabic{equation}}
\setcounter{equation}{0}
\section*{Appendix A}

The contribution of the diagram in fig.~3 for the $\d (1232)$
resonance to the amplitude of the process under consideration can be written
as
\be\label{del}
T_{\d}&=&\frac{-2e^2(\e_2q_2)g_0(s)/\mm}{(s_1-\mm^2)(s-M^2_{\d}+
iM_{\d}\Gamma(s))}\bar u(p_2)q^{\m}\d_{\m\n}(p)\g_5  \\
& &\times\left[-(g_1(s)/m)(k_1^{\n}\hat \e_1-\e_{1}^{\n}\hat k_1)+
(g_2(s)/m^2)(k_{1}^{\n}(p_1\e_1)-\e_{1}^{\n}(p_1k_1))\right]u(p_1),\nonumber
\en
where
\EQ\label{prop}
\d_{\m\n}(p)=(1/3M^2_{\d})(\hat p+M_{\d})(2p_{\m}p_{\n}-3M^2_{\d}g_{\m\n}+
M^2_{\d}\g_{\m}\g_{\n}+M_{\d}(\g_{\m}p_{\n}-\g_{\n}p_{\m})),
\EN
$p=p_1+k_1$, $q=q_2+k_2$ and $\Gamma$ is the decay width of the $\d(1232)$
resonance.

In accordance with refs. \cite{blom,bar} the following parametrization is
used:
$$
\Gamma(s)=\Gamma_0\left(\frac{|q|}{|q_{\d}|}\right)^3\frac{M_{d}}{\sqrt{s}}
\frac{1+R^2|q_{\d}|^2}{1+R^2|q|^2},
$$
$$
g_0(s)=g_0((1+R^2|q_{\d}|^2)/(1+R^2|q|^2))^{1/2}, \qquad
|q_{\d}|=|q(s=M^2_{\d})|,
$$
$$
g_{1,2}(s)=g_{1,2}((1+R^2|k_{\d}|^2)/(1+R^2|k|^2))^{1/2}, \qquad
|k_{\d}|=|k(s=M^2_{\d})|,
$$
\begin{center}
$M_{\d}=1232$ MeV, \quad $\Gamma_0=109$ MeV, \quad $R=5.5$ GeV$^{-1}$.
\end{center}

According to ref. \cite{bar} the coupling constants are taken to be
$g_0g_1/\mm m=19.78$ GeV$^{-2}$ and $g_0g_2/\mm m^2=-21.2$ GeV$^{-3}$.

Since the term at the coupling $g_2$ in eq. (\ref{del}) gives a major 
contribution to the
$E_{1+}$ electric quadrupole \cite{blom}, we represent the contribution of
the $D_{13}$ resonance as
\be\label{D}
T_{D_{13}}&=&\frac{-2e^2(\e_2q_2)\bar g_0(s)/\mm}{(s_1-\mm^2)(s-M^2_D+
iM_D\bar \Gamma(s))}\bar u(p_2)q^{\m}\g_5\bar \d_{\m\n}(p) \nonumber\\
& &\times(\bar g_2(s)/m^2)[k_{1}^{\n}(p_1\e_1)-\e_{1}^{\n}(p_1k_1)]u(p_1).
\en

The propagator $\bar \d_{\m\n}(p)$ is obtained from $\d_{\m\n}(p)$
by replacing the mass $M_{\d}$ by $M_D$. We use the same parametrization
as for the $\d(1232)$ resonance but with
the following values of the mass, width, and coupling constants:
$M_D=1520$ MeV, $\bar g_0 \bar g_2=0.5 g_0g_2$, $\bar
\Gamma_0=\Gamma_0=109$~MeV.

The contribution of the $P_{11}(1440)$ resonance can be written as \cite{bar}
\be\label{rop}
T_{P_{11}}&=&\frac{-2e^2(\e_2q_2)g_0^*}{(s_1-\mm^2)(s-M^2_{P}+iM_{P}
\Gamma_P(s))}
\bar u(p_2)\g_5(\hat p_1+\hat k_1+M_P)  \nonumber\\
&& \left[(g_1^*/2M_P)\hat k_1\hat\e_1+(g_2^*/M^2_P)\left((p_1\e_1)
\hat k_1-(p_1k_1)\hat\e_1\right)\right]u(p_1),
\en
where $M_P$ is the mass of the $P_{11}(1440)$ resonance,
$$
\Gamma_P(s)=\Gamma_{P0}\frac{(s-m^2)M_P}{(M^2_P-m^2)\sqrt{s}},
$$
and with the following values of mass, width, and
coupling constants \cite{bar}: $M_P=1440$ MeV,
$\Gamma_{P0}=200$ MeV, $g^*_0g^*_1/2M_P=-2.7$ GeV$^{-1}$,
and $g^*_0g^*_2/M_P^2=-21.3$ GeV$^{-2}$.

The amplitude of the $S_{11}(1535)$ resonance is taken as
\EQ\label{s11}
T_{S_{11}}=\frac{-g_{1S}g_{2S}(\e_2q_2)}{(s_1-\mm^2)(s-M^2_S+iM_S\Gamma_S(s))}
\bar u(p_2)\left[\hat p_2+\hat q +M_S\right]\g_5 \hat\e_1 u(p_1),
\EN
where $M_S$ is the mass of the $S_{11}(1535)$ resonance and
$$
g^2_{1S}=\frac{8\pi\Gamma_{S\to\g p}M_S^3}{(M^2_S-m^2)(M^2_S+m^2+4mM_S)},
$$
$$
g^2_{2S}=\frac{8\pi\Gamma_{S\to\pi n}M^2_S}{q(M_S)[(M_S+m)^2-\mm^2]},
$$
with $M_S$=1535 MeV, $\Gamma_S$=150 MeV, $\Gamma_{S\to\g p}$=375 MeV,
and $\Gamma_{S\to\pi n}$=75 MeV.

However, the considered amplitudes of the nucleon resonances are not gauge
invariant and
besides the calculations with them are cumbersome. These problems are
solved by using the method of orthogonal amplitudes \cite{pr,hearn,bar1}.

\renewcommand{\theequation}{B\arabic{equation}}
\setcounter{equation}{0}
\section*{Appendix B}


Let us choose the
following basis of orthogonal vectors
\be\label{bas}
K&=&k_1+k_2,\nonumber\\
Q&=&k_1-k_2,\nonumber\\
P^{\i}&=&P-\frac{(PK)}{K^2}K-\frac{(PQ)}{Q^2}Q,\\
N_{\m}&=&\e_{\m\n\lambda\s} P^{\i\n}K^{\lambda}Q^{\s},\nonumber
\en
where $P=p_1+p_2$ and $\e_{0123}=+1$

Let us write the amplitude of the process as
\EQ
\bar u(p_2)Tu(p_1)=\bar u(p_2)\e_{2}^{\m}T_{\m\n}\e_{1}^{\nu}u(p_1)
\EN
and expand $T_{\m\n}$ in terms of the basis vectors $\eta_{\m}^{\s}$
\EQ
T_{\m\n}=\sum_{\s\s\i} \eta^{\s}_{\m}C_{\s\s\i}\eta^{\s\i}_{\n}.
\EN
Eight products can be constructed from the vectors $\e_2$, $\e_1$, $K$, $Q$,
$P^{\i}$, and $N$:
$$
(\e_2K),\quad (\e_1K),\quad (\e_2Q),\quad (\e_1Q),\quad (\e_2P^{\i}),\quad
(\e_1P^{\i}),\quad
(\e_2N),\quad (\e_1N).
$$
To reduce the number of possible combinations, we use the gauge
invariance conditions:
\EQ
(\e_2k_2)=(\e_1k_1)=0,
\EN
\EQ\label{gin}
T_{\m\n}k_{1}^{\n}=k_{2}^{\m}T_{\m\n}=0.
\EN
Then it is evident from (\ref{gin}) that
$$
T_{\m\n}(K-Q)^{\n}=0 \quad {\rm or} \quad T_{\m\n}K^{\n}=T_{\m\n}Q^{\n}.
$$
If $\eta_{\n}=K_{\n}$, then
$$
K^2=(KQ)=0,
$$
and similarly, if $\eta_{\n}=Q_{\n}$, then
$$
(QK)=Q^2=0.
$$
However, $K^2$ and $Q^2$ do not vanish identically. Therefore, the
products $(\e_i K)$ and $(\e_i Q)$ are forbidden and as a consequence,
only the following combinations remain:
$$
(\e_2P^{\i})(\e_1P^{\i}),\quad (\e_2N)(\e_1N),
$$
\EQ\label{comb}
(\e_2P^{\i})(\e_1N)\pm (\e_2N)(\e_1P^{\i}).
\EN
It follows from parity conservation that
the coefficients of the first two terms are scalars, and those of the last
two are pseudoscalars.

To consider the Dirac structure of the operator $C_{\s\s^\i}$, we
construct the following set of orthogonal vectors:
\be\label{spin}
P&=&(p_1+p_2), \nonumber\\
L&=&(p_1-p_2), \nonumber\\
V&=&K-\frac{(KP)}{P^2}P-\frac{(KL)}{L^2}L, \\
R_{\m}&=&\e_{\m\n\lambda\s} V^{\n}P^{\lambda}L^{\s}. \nonumber
\en

The quantities $\hat P$ and $\hat L$ are effectively $c$--numbers by virtue
of the Dirac equation and the commutation relations.
An additional $\hat R$ and $\hat R\hat V$ may be eliminated since they
can be expressed through $\g_5\hat V$ and $\g_5$.

In conclusion,
the gauge invariant amplitude of the process $\gp$ may be written in the
orthogonal basis as
\be\label{ampl}
\lefteqn{\bar u(p_2)Tu(p_1)=\bar u(p_2)\g_5\left\{\frac{(\e_2P^{\i})
(\e_1P^{\i})}{(P^{\i})^2}\left[T_1+\hat VT_2\right]\right.} \nonumber\\
& &+\frac{(\e_2N)(\e_1N)}{N^2}\left[T_3+\hat V T_4\right] \nonumber\\
& &+\frac{(\e_2P^{\i})(\e_1N)-(\e_2N)(\e_1P^{\i})}{\sqrt{(P^{\i})^2N^2}}
 i\g_5\left[T_5+\hat VT_6\right] \\
& &+\left.\frac{(\e_2P^{\i})(\e_1N)+(\e_2N)(\e_1P^{\i})}{\sqrt{(P^{\i})^2N^2}}
 i\g_5\left[T_7+\hat VT_8\right]\right\}u(p_1). \nonumber
\en

As a result, the differential cross section of the process under consideration
is
\be\label{rcr}
&&\frac{d\s_{\gp}}{dt ds_{1} d\O_{\g\g}^{cm}}=\frac{1}{64\pi^2}
(\frac{e^2}{4\pi})^2\frac{(s_1-\mm^2)}{s_1(s-m^2)^2} \nonumber\\
&&\times \left\{-t\left(|T_1|^2+|T_3|^2\right)
-V^2P^2\left(|T_2|^2+|T_4|^2\right)\right. \\
&&\left.+2P^2\left(|T_5|^2+|T_7|^2\right)
+2V^2 t\left(|T_6|^2+|T_8|^2\right)\right\}. \nonumber
\en

An obvious advantage of this method is that the amplitude (\ref{ampl}) is
gauge invariant. Moreover, in this case the amount of calculations is
proportional to the number $N$ of diagrams considered and not to $N^2$ which
is usually the case in ordinary calculations of the cross section.

\renewcommand{\theequation}{C\arabic{equation}}
\setcounter{equation}{0}
\section*{Appendix C}

In this appendix we present the method of a projection of amplitudes of any
diagram on the scalar amplitudes $T_i$ of eq. (\ref{ampl}).
With this aim we expand the amplitude for the diagram under consideration
in terms of the full set of the basis vectors (\ref{bas}),
\EQ
\bar u(p_2)A_{\m\n}u(p_1)=\bar u(p_2)\sum_{\s\s^{\i}}\eta_{\m}^{\s}
C_{\s\s^{\i}}\eta_{\n}^{\s^{\i}}u(p_1).
\EN
Taking account of the orthogonality of the vectors, we obtain the
coefficients of the expansion as
\EQ\label{coef}
\bar u(p_2)C_{\s\s^{\i}} u(p_1)=\bar u(p_2)\frac{\eta^{\s\m}A_{\m\n}
\eta^{\s^{\i}\n}}{(\eta^{\s})^2(\eta^{\s^{\i}})^2}u(p_1).
\EN
The functions $C_{\s\s^{\i}}$ may depend on the matrices $\g_5$, $\hat V$,
and $\g_5\hat V$. In order to determine the scalar amplitudes $T_i$, we
multiply eq.~(\ref{coef})
from the left by $u(p_2)$ and from the right by $\bar u(p_1)$
and use the relation $u(p)\bar u(p)=(\hat p+m)/2m$. Then, multiplying
the left and the right sides of the equation by the operator $O$
($O=1$, $\g_5$, $\hat V$, $\g_5\hat V$)  and
calculating the traces of these expressions, we find a set of linear
equations that determine the $T_i$ of eq. (\ref{ampl}),
\EQ
Tr\left\{ O(\hat p_2+m)C_{\s\s^{\i}}(\hat p_1+m)\right\}=
\frac{1}{(\eta^{\s})^2(\eta^{\s^{\i}})^2} Tr\left\{ O(\hat
p_2+m)\eta^{\s\m}
A_{\m\n}\eta^{\s^{\i}\n}(\hat p_1+m)\right\}.
\EN

\newpage
\section*{ Table captions}
\begin{enumerate}
\item The experimental data presently available for the pion
polarizabilities.
\item Test run results for counting rates measurements of each
MWPC plane at 8 mm collimator, 3.25~kV MWPC high voltage and for
a tagged interval of $520-795$ MeV.
\end{enumerate}

\newpage
\section*{Figure captions}

\begin{enumerate}
\item The diagram for the radiative pion photoproduction from the
proton.
\item The nucleon and pion pole diagrams in the model with
pseudoscalar coupling.
\item The nucleon and pion pole diagrams as well nucleon resonance
contributions to radiative $\pp$ photoproduction from the proton.
\item Floor plan of the experimental setup showing the location of
the detectors. A, B, C are TAPS blocks, MWPC+FSD show multi-wire
proportional chambers and the forward scintillation detector,
TOF indicates the block of the neutron detector bars, and LH$_2$
stands for the liquid hydrogen target in its vacuum scattering
chamber.
\item Enlarged view showing the details of the TAPS configuration.
\item The arrangement of the planes of the multi-wire proportional
chambers.
\item The angular dependence of the differential cross section for
$\g p\to\pn p$ in the energy range $480-530$ MeV. The open circles are
the data from ref. \cite{mami}, the filled circles are the data of the present
work. The solid, dashed, and dotted lines are results of the MAID, DMT,
and SAID analysis, respectively.
\item A typical triple coincidence time spectrum taken after the
kinematical cuts;
$t_{\g}$ is defined by the electron ladder in the tagger, $t_{\g^{\i}}$
is given by TAPS, and $t_{\pi}$ by the forward hodoscope.
\item The missing-mass spectrum of the reactions $\g p\to\g nX$ taken
after applying the kinematical and time cuts and subtraction of the random
background.
The solid line is the experimental spectrum and the dashed line is
the result of the simulation.
\item The differential cross section of the process $\gp$
averaged over the full photon beam energy interval and over $s_1$
from $1.5\mm^2$ to $5\mm^2$. The solid and dashed lines
are the predictions of model-1 and model-2, respectively,
for $(\a-\b)_{\pp}=0$. The dotted line is a fit to the
experimental data (see text).
\item The cross section of the process $\gp$
integrated over $s_1$ and $t$ in the region where the contribution of the
pion polarizability is biggest and the difference between the predictions
of the theoretical models under consideration does not exceed 3\%.
The dashed and dashed-dotted lines are predictions of model-1
and the solid and dotted lines of model-2
for $(\a-\b)_{\pp}=0$ and $14 \unit$, respectively.
\item The dependence of the cross section $\s$ for $\gp$
on $(\a-\b)_{\pp}$ at $E_\gamma = 653$ MeV as obtained in the framework
of model-1. The experimental values of the cross section are given
with their statistical and systematic errors.
\end{enumerate}

\newpage
\begin{table}       
\centering
\begin{tabular}{|ll|c|c|} \hline
\multicolumn{2}{|l|}{Experiments} & $\a_{\pi^{\pm}}\Unit$ &
$\a_{\pn}\Unit$  \\ \hline
\multicolumn{2}{|l|}{$\pi^{-}Z\rightarrow\g \pi^{-} Z$, Serpukhov (1983)
\cite{antip}} & $6.8\pm 1.4\pm 1.2$ &     \\  \hline
\multicolumn{2}{|l|}{$\gp$, Lebedev Phys.Inst. (1984) \cite{lebed}} &
$20\pm 12$ &     \\  \hline
\multicolumn{2}{|l|}{D. Babusci {\em et al.} (1992) \cite{bab}} &   &   \\
$\g \g\rightarrow \pp \pi^{-}$:   & PLUTO (1984) \cite{pluto} &
$19.1\pm 4.8\pm 5.7$ &      \\
    & DM 1 (1986) \cite{dm1} & $17.2\pm 4.6$ &     \\
    & DM 2 (1986) \cite{dm2} & $26.3\pm 7.4$ &     \\
    & MAPK II (1990) \cite{mark} & $2.2\pm 1.6$ &    \\
$\g \g\to \pn \pn$:  & Crystal Ball (1990) \cite{crb} &    &
$ \pm 0.69\pm 0.11$  \\  \hline
\multicolumn{2}{|l|}{F. Donoghue, B. Holstein (1993) \cite{holst}} &   &   \\
$\g \g\to \pp \pi^{-}$:   & MARK II  & $2.7$     &   \\
$\g \g\to \pn \pn$:  & Crystal Ball  &    & $-0.5$  \\ \hline
    &     &$(\a+\b)_{\pn}\Unit$ & $(\a-\b)_{\pn}\Unit$   \\ \hline
\multicolumn{2}{|l|}{A. Kaloshin, V. Serebryakov (1994) \cite{ser}
} &   &   \\
$\g \g\to \pn \pn$:  & Crystal Ball  &$1.00\pm 0.05$ & $-0.6\pm 1.8$
\\ \hline
\multicolumn{2}{|l|}{L. Fil'kov, V. Kashevarov (1999) \cite{kash}}
&   &   \\
$\g \g\to \pn \pn$:  & Crystal Ball  &$0.98\pm 0.03$ & $-1.6\pm 2.2$
\\ \hline
\end{tabular}
\caption{}
\end{table}

\newpage

\begin{table}     
\centering
\begin{tabular}{|c|c|c|c|c|c|c|c|} \hline
MWPC & BEAM & MWPC1 & MWPC2 & singl & max. & FSD & max. \\
plane & current & current & current & plane & wire & rate & strip \\
number & nA & $\mu$A & $\mu$A & rate & rate & MHz & rate \\
 &  &  &  & MHz & MHz &  & MHz \\
\hline
1 & 155.4 & 75.0 & 73.0 & 7.78 & 0.159 & 9.59 & 1.44 \\
\hline
2 & 155.4 & 75.0 & 73.0 & 6.09 & 0.137 & 9.53 & 1.41 \\
\hline
3 & 158.7 & 76.0 & 74.0 & 6.35 & 0.176 & 9.64 & 1.42 \\
\hline
4 & 155.6 & 75.0 & 73.0 & 7.63 & 0.199 & 9.44 & 1.40 \\
\hline
\end{tabular}
\caption{}
\end{table}

\newpage
\begin{figure}\label{diag} 
\epsfxsize=8cm
\epsfysize=8cm
\centerline{
\epsffile{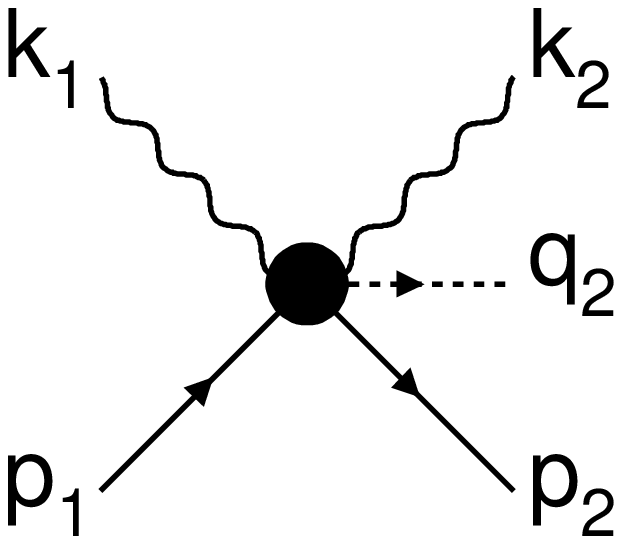}}
\caption{}
\end{figure}

\newpage
\begin{figure}\label{psc} 
\epsfxsize=17cm
\epsfysize=12cm
\centerline{
\epsffile{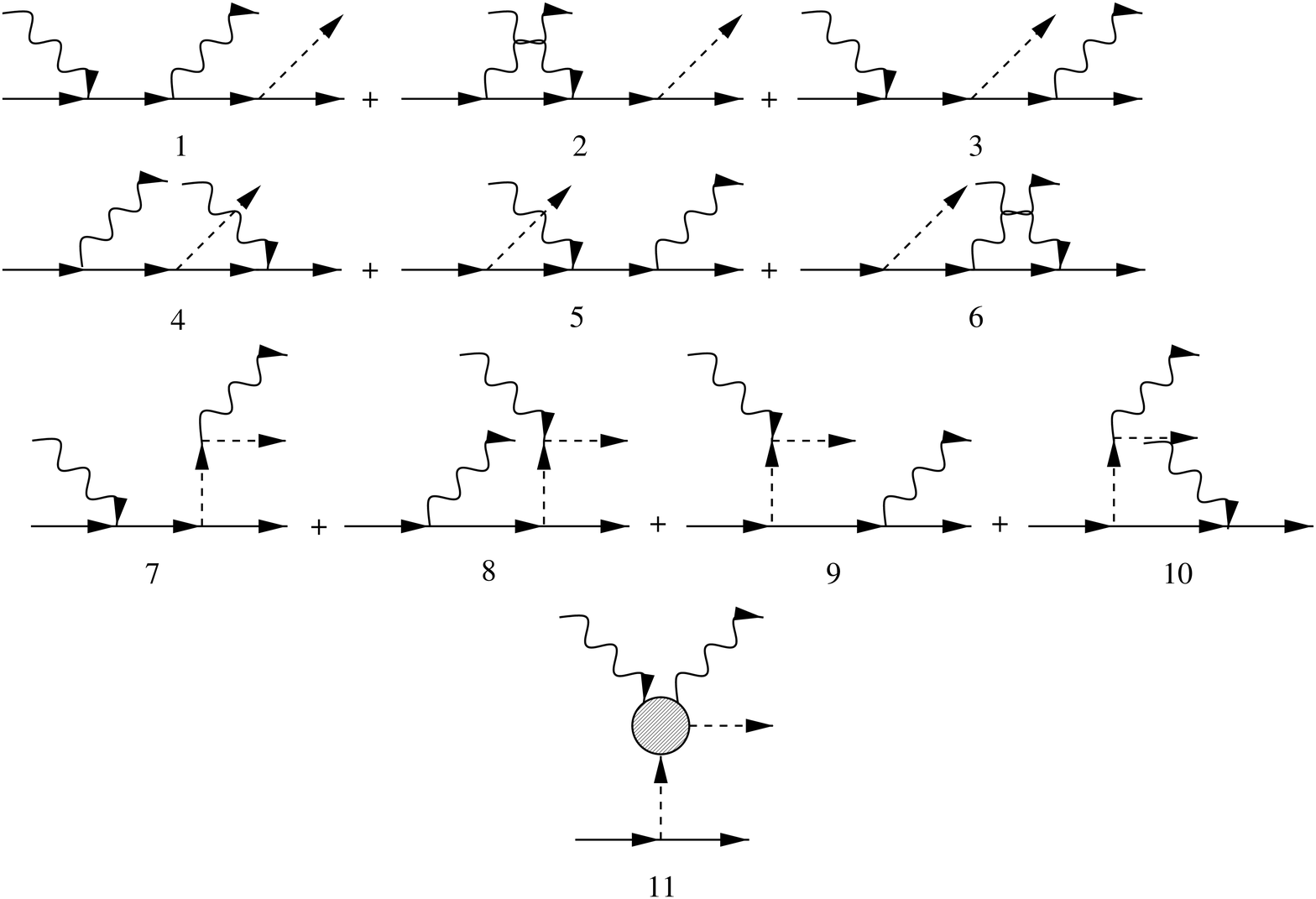}}
\caption{}
\end{figure}

\newpage
\begin{figure}[H]\label{diag2} 
\epsfxsize=17cm
\epsfysize=8cm
\centerline{
\epsffile{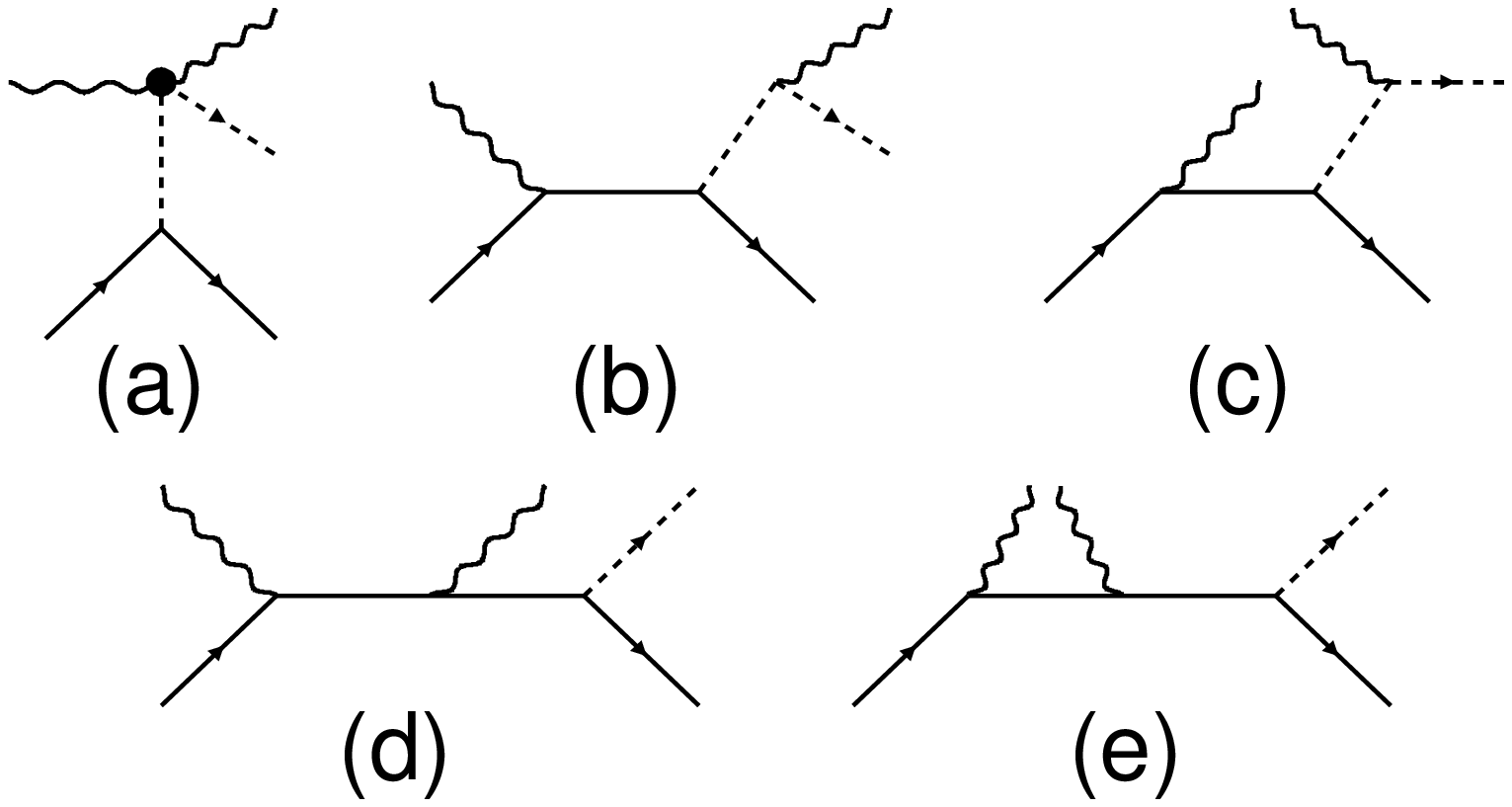}}
\epsfxsize=9cm
\epsfysize=6cm
\centerline{
\epsffile{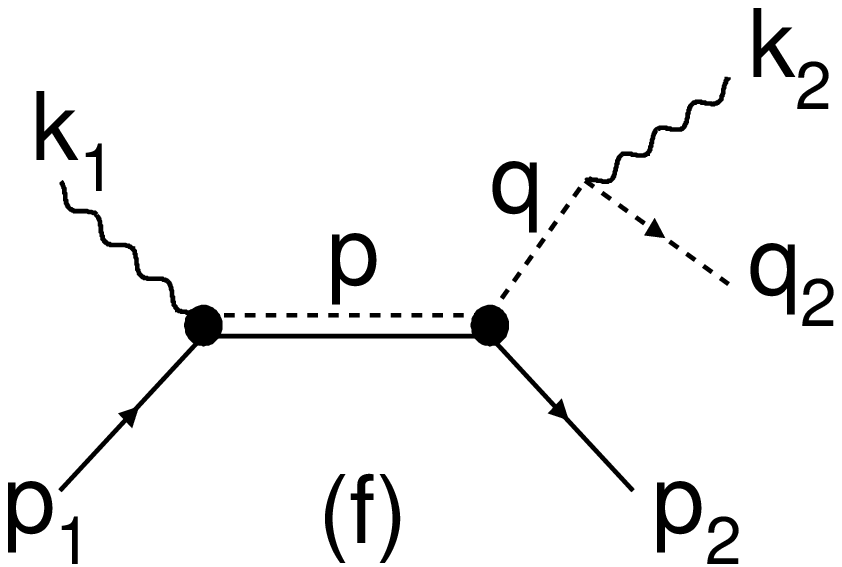}}
\caption{}
\end{figure}

\newpage
\begin{figure}\label{setup}  
\epsfxsize=12cm
\epsfysize=12cm
\centerline{
\epsffile{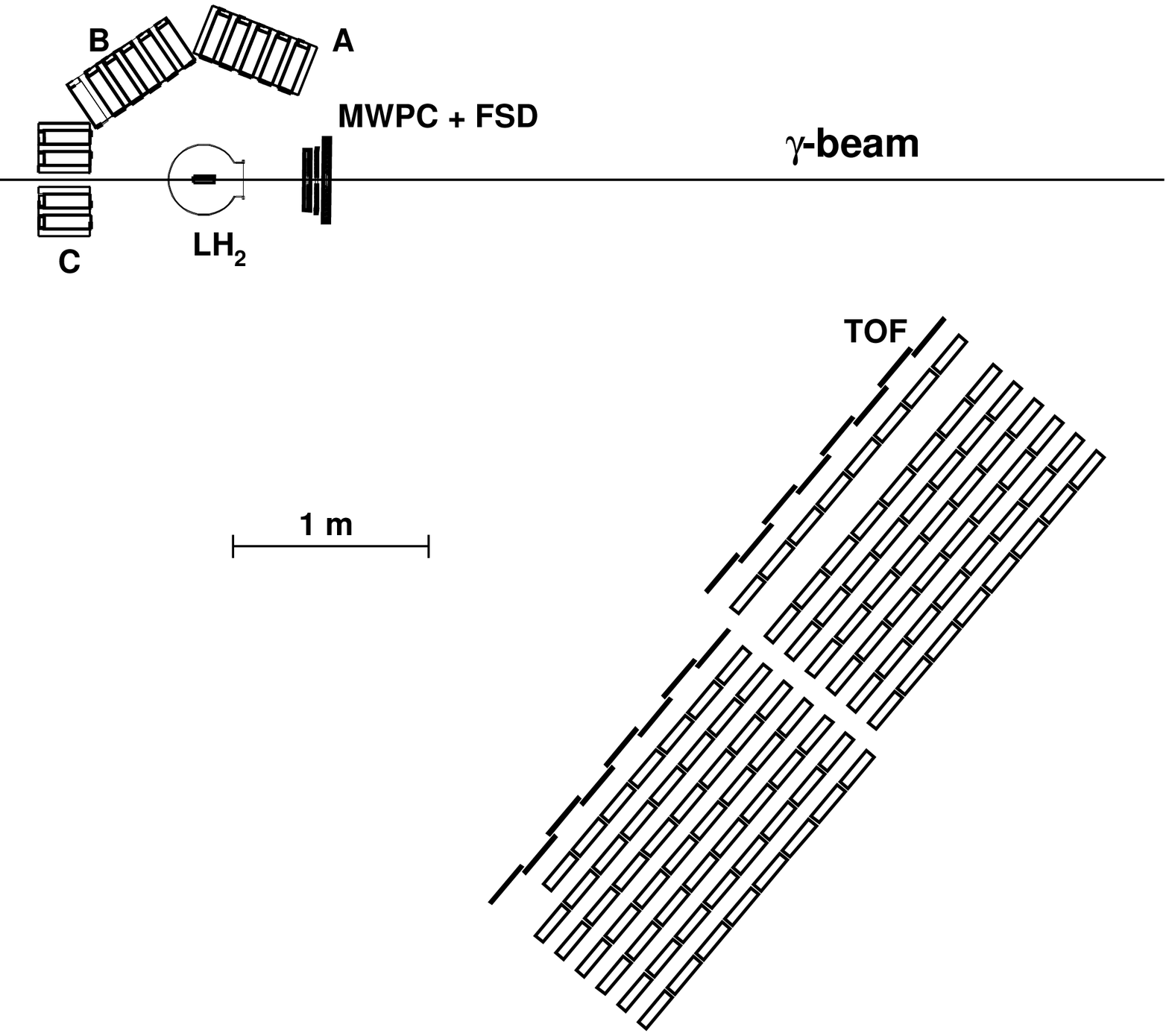}}    
\caption{}
\end{figure}

\newpage
\begin{figure}\label{taps} 
\epsfxsize=10cm
\epsfysize=10cm
\centerline{
\epsffile{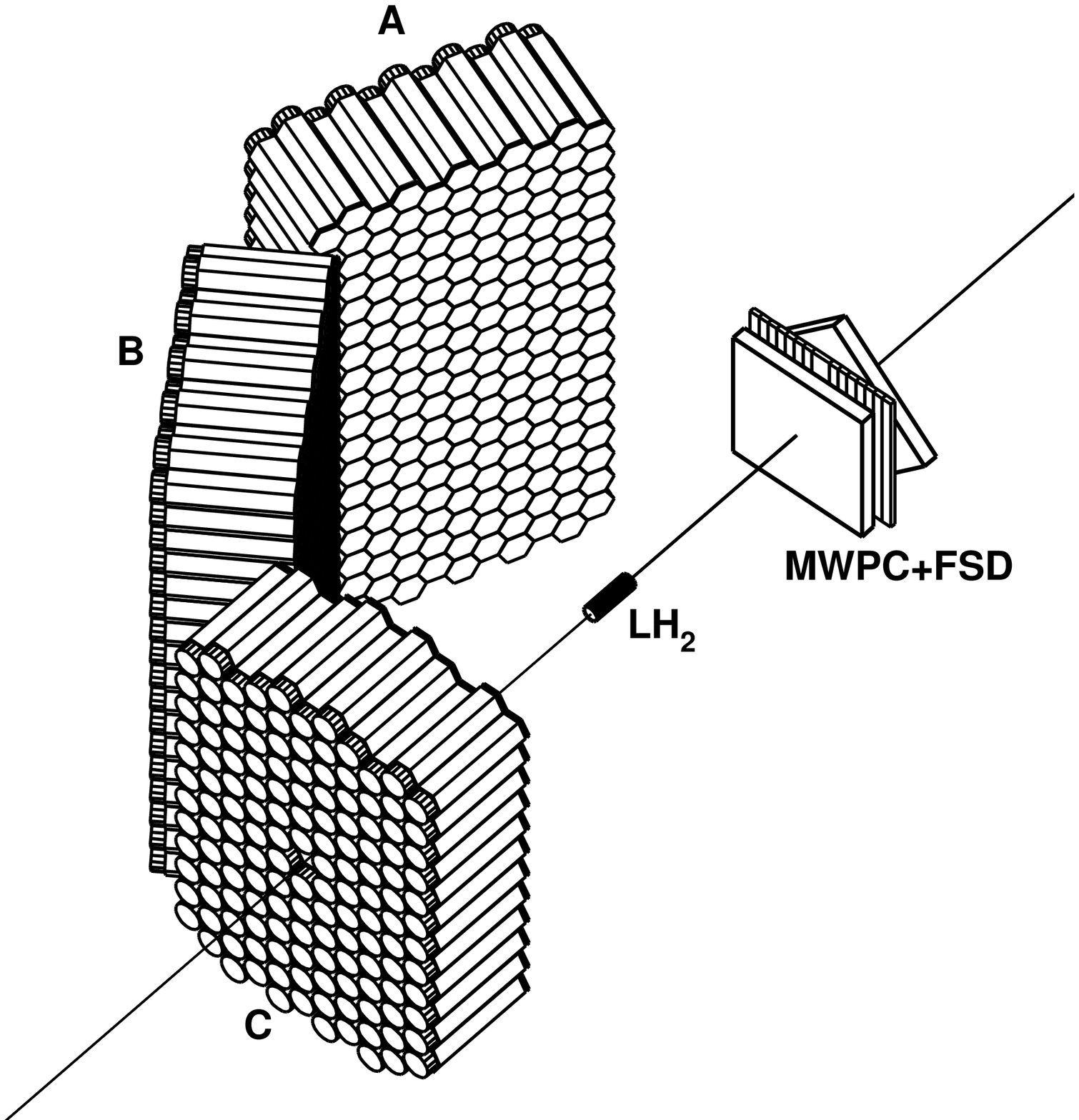}}    
\caption{}
\end{figure}

\newpage

\begin{figure}\label{mwpc} 
\epsfxsize=10cm
\epsfysize=10cm
\centerline{
\epsffile{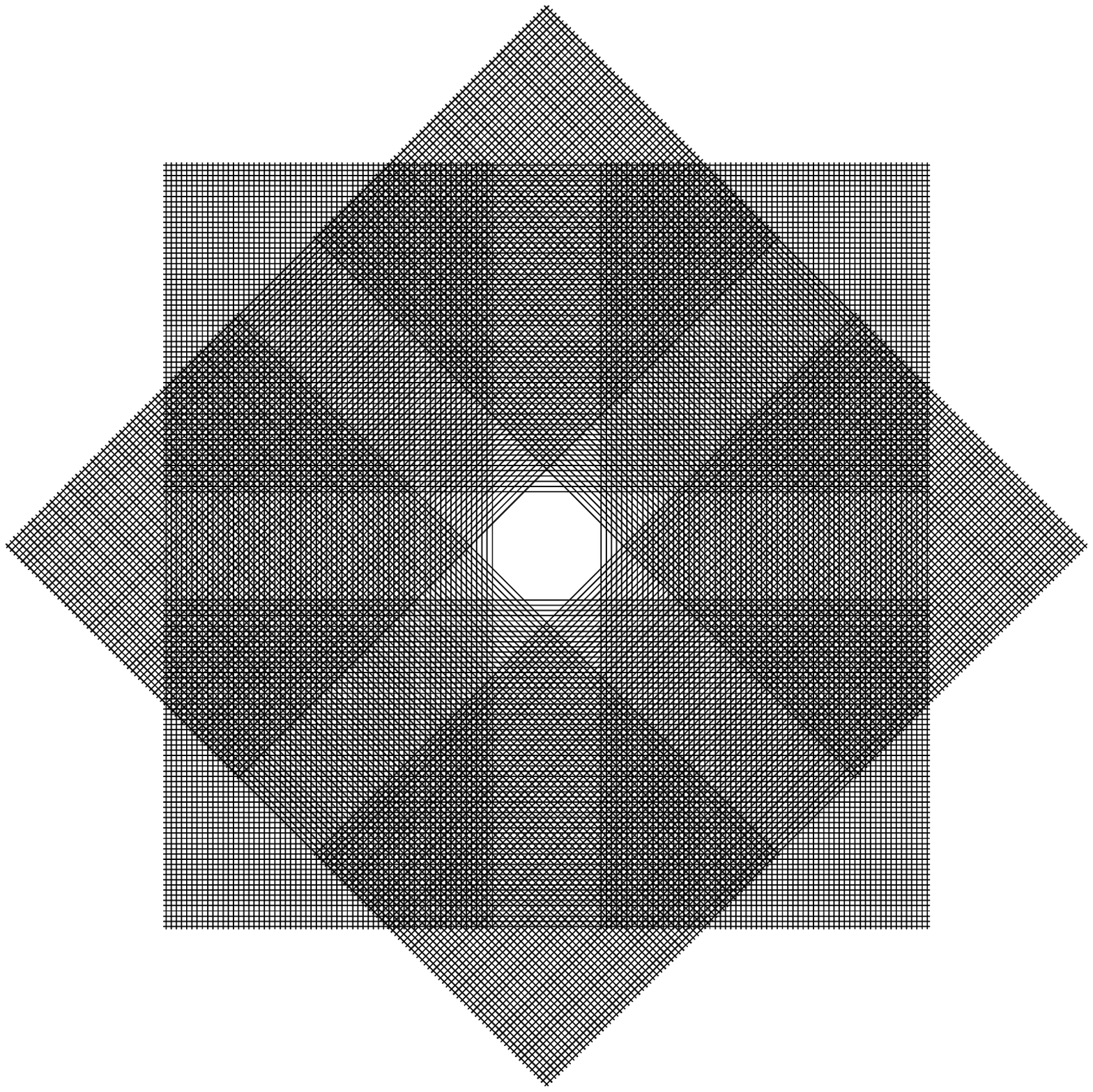}}
\caption{}
\end{figure}

\newpage
\begin{figure}\label{pi0}  
\epsfxsize=15cm
\epsfysize=20cm
\centerline{
\epsffile{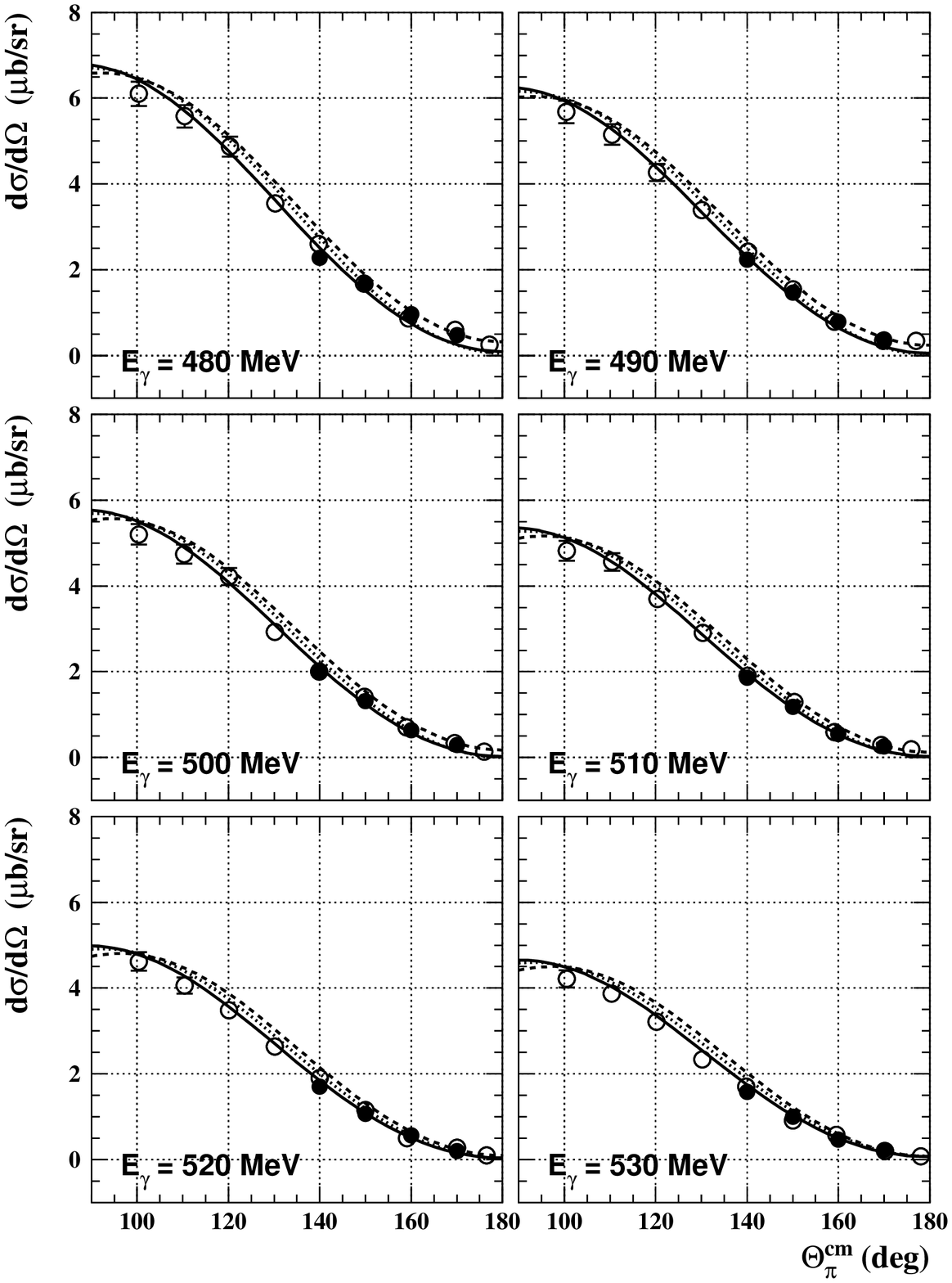}}
\caption{}
\end{figure}

\newpage
\begin{figure}\label{time}  
\epsfxsize=10cm
\epsfysize=12cm
\centerline{
\epsffile{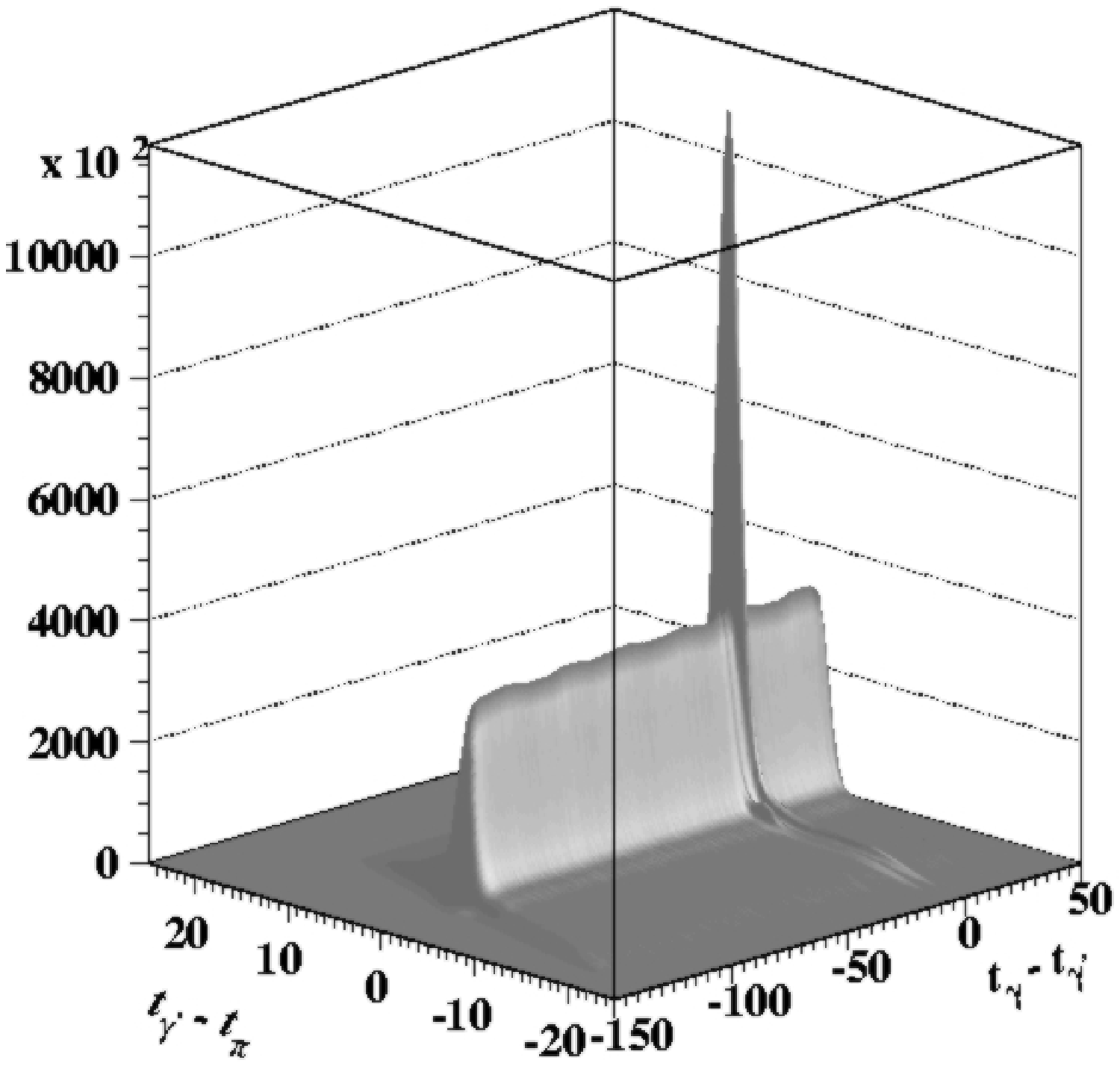}}   
\caption{}
\end{figure}

\newpage
\begin{figure}\label{mis}  
\epsfxsize=10cm
\epsfysize=12cm
\centerline{
\epsffile{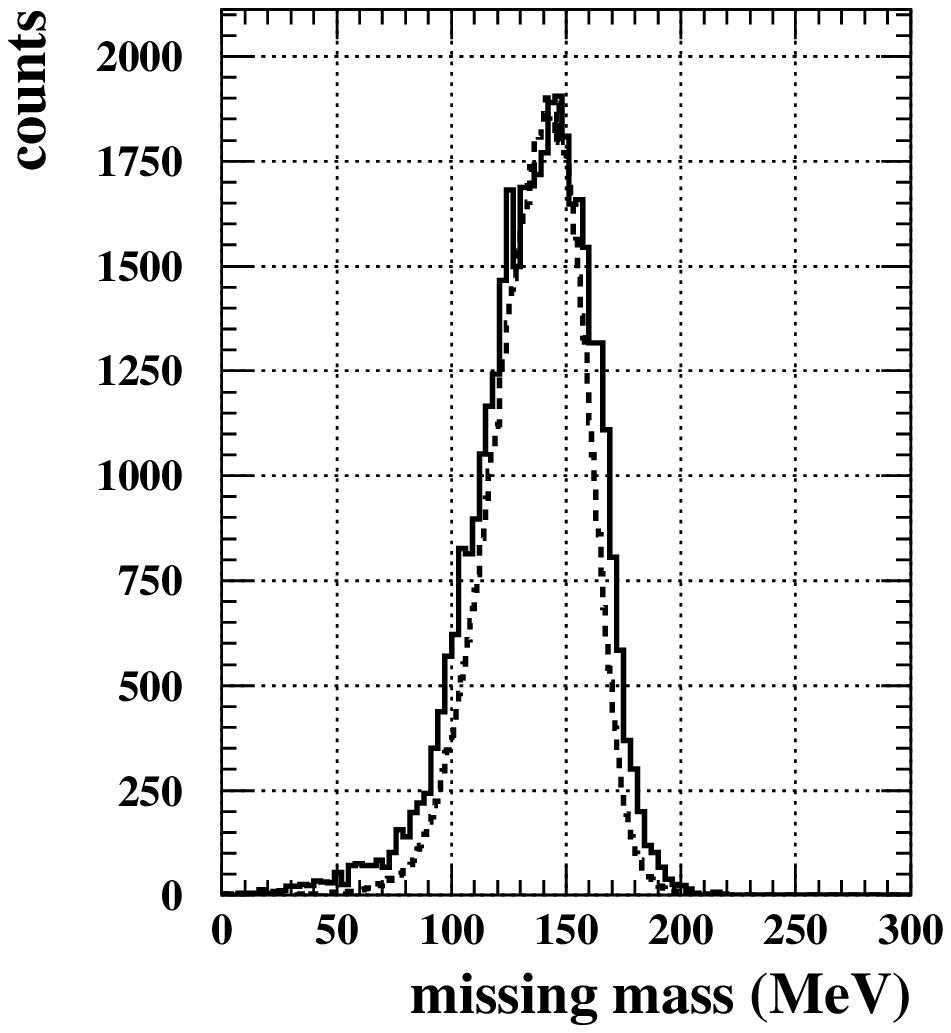}}
\caption{}
\end{figure}

\newpage
\begin{figure}\label{cros1} 
\epsfxsize=15cm
\epsfysize=11cm
\centerline{
\epsffile{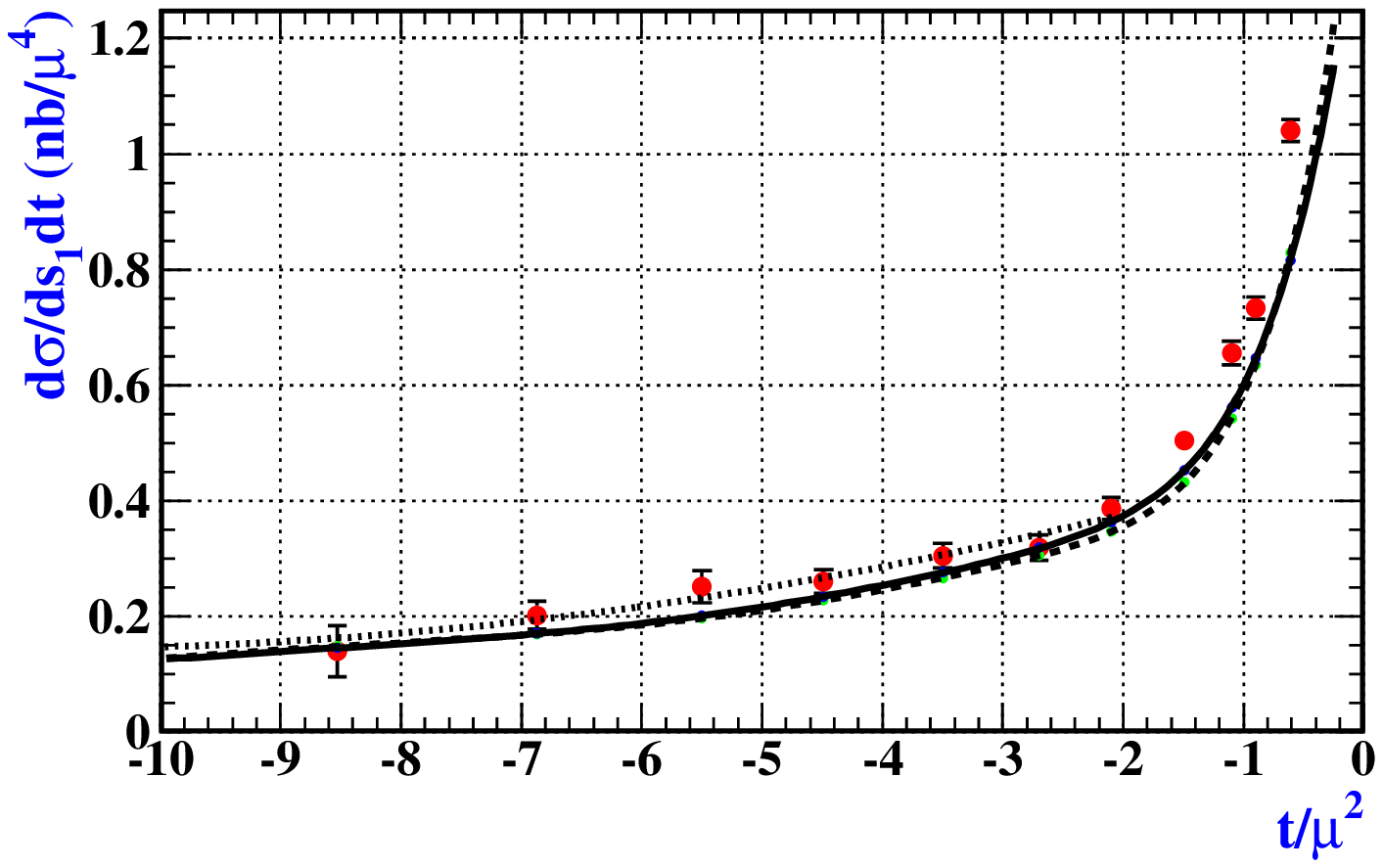}}   
\caption{}

\end{figure}

\newpage
\begin{figure}\label{cros2} 
\epsfxsize=15cm
\epsfysize=11cm
\centerline{
\epsffile{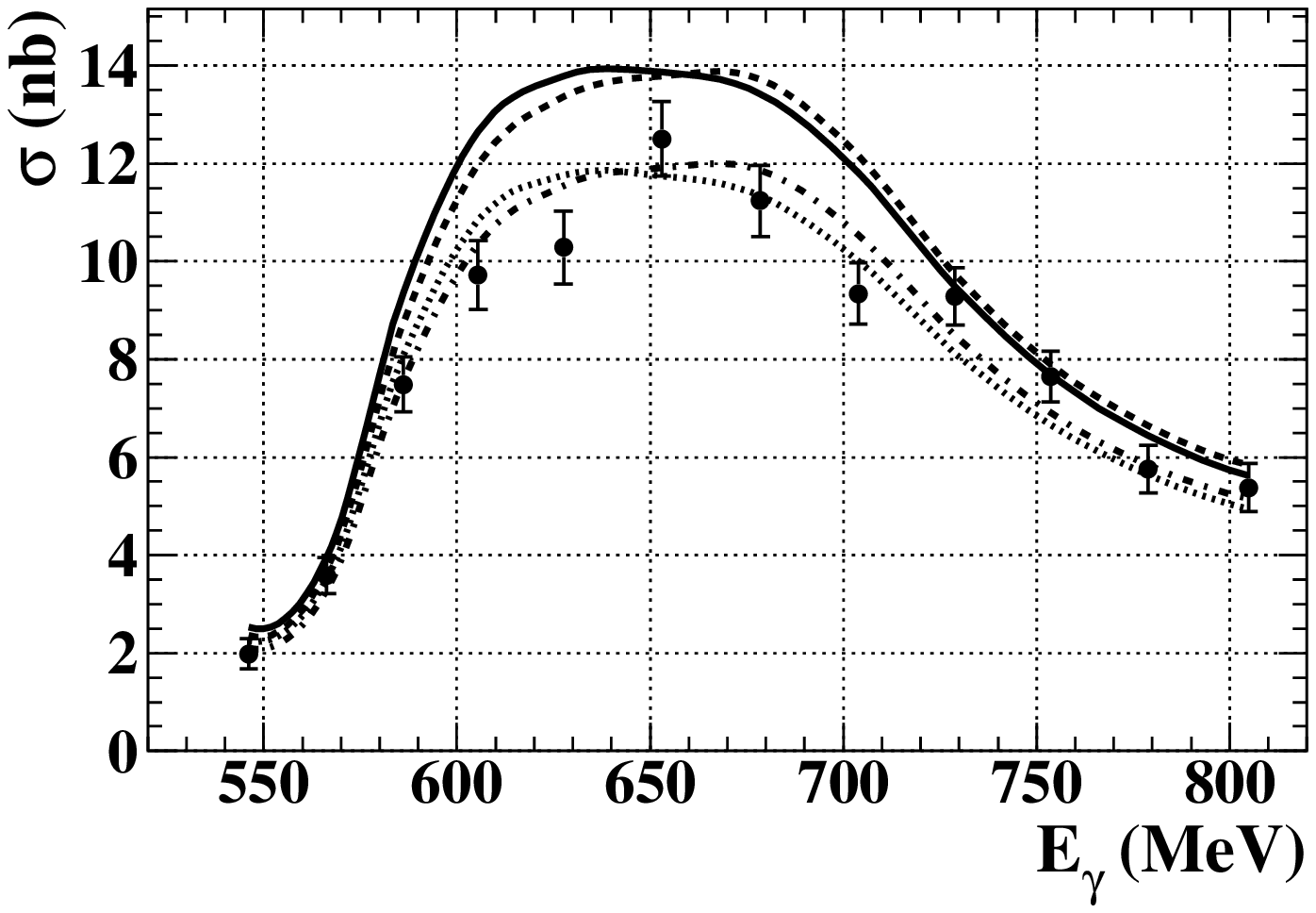}}   
\caption{}
\end{figure}

\newpage
\begin{figure}\label{cros3} 
\epsfxsize=10cm
\epsfysize=12cm
\centerline{
\epsffile{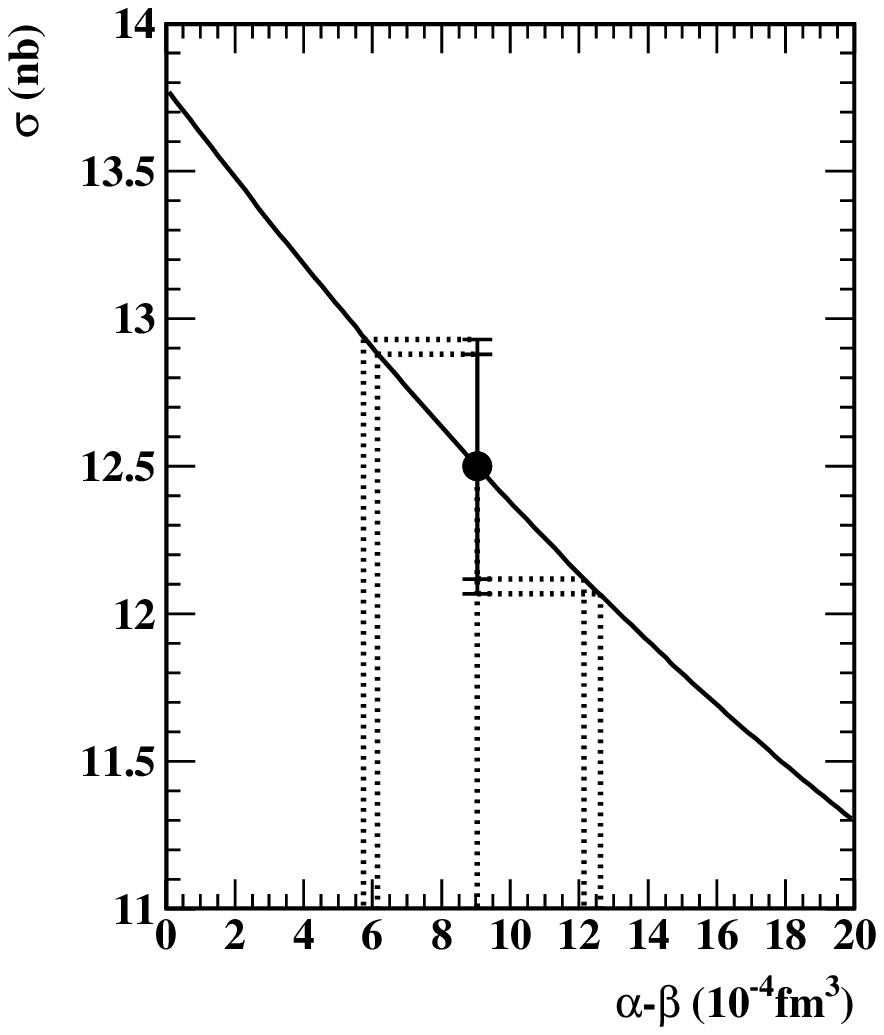}}   
\caption{}
\end{figure}

\end{document}